\documentclass[a4paper,11pt]{article}
\pdfoutput=1 

\usepackage{jcappub} 

\usepackage[T1]{fontenc} 
\usepackage{footnote}

\usepackage{graphicx,ulem}
\usepackage{longtable}
\usepackage{float}
\usepackage{dcolumn}
\usepackage{graphics,epsfig}
\usepackage{amsmath,amssymb,latexsym,mathrsfs}
\usepackage{bm}
\usepackage{color}
\usepackage{color}
\usepackage{subfigure}
\usepackage{multirow}
\usepackage{amsfonts}

\begin{document}

\title{ Testing Predictions of the Quantum Landscape Multiverse 1: The Starobinsky Inflationary Potential}

\author[1,2]{Eleonora Di Valentino}
\author[3]{and Laura Mersini-Houghton} 

\affiliation[1]{Institut d' Astrophysique de Paris (UMR7095: CNRS \& UPMC-Sorbonne Universities), F-75014, Paris, France}
\affiliation[2]{Sorbonne Universit\'es, Institut Lagrange de Paris (ILP), F-75014, Paris, France}
\affiliation[3]{Department of Physics and Astronomy,
    UNC-Chapel Hill, NC 27599, USA}

\date{\today}

\abstract{The 2015 Planck data release has placed tight constraints on the allowed class of inflationary models. The current data favors concave downwards inflationary potentials while offering interesting hints on possible deviations from the standard picture of CMB perturbations. We here test the predictions of the theory of the origin of the universe from the landscape multiverse, against the most recent Planck data, for the case of concave downwards inflationary potentials, such as the Starobinsky model of inflation. By considering the quantum entanglement correction of the multiverse, we can place a lower limit on the local 'SUSY breaking' scale $b>1.2\times10^7 GeV$ at $95 \%$ c.l. from Planck TT+lowTEB. We find that this limit is consistent with the range for $b$ that allows the landscape multiverse to explain a serie of anomalies present in the current data.}


\maketitle
\flushbottom

\section{Introduction}

Recent successes in precision cosmology place us in a position where we can discriminate among various inflationary models, while for the first time test new physics beyond the standard model of cosmology. The latest evidence on the existence of anomalies accumulated by the Planck collaboration data \cite{planck1,planck2,planck2013} provides enticing probes into the pre-inflationary universe.

In this work we compare the predictions of the theory of the origin of the universe from a quantum landscape multiverse \cite{archillmh,richlmh,tomolmh,lmh} against the recent Planck data. Based on a 7 parameter likelihood exploration of standard cosmology, the recent results from the Planck collaboration data tightly constrain the tensor-to-scalar ratio $r$ in the minimal standard cosmological model. As a result, from the plethora of models, the current data favor pure LCDM or concave downward inflationary potentials, such as the Starobinsky model \cite{staro} and the class of hilltop inflationary models \cite{hilltop}. Most of the convex potentials resulting in large $r$ seem to be ruled out. 

Modifications to the perturbation spectrum and the gravitational potential, originating from quantum entanglement in the theory of the origin of the universe from the quantum multiverse \cite{archillmh,richlmh}, were derived in \cite{lmh} for concave downward inflationary potentials and in \cite{tomolmh} for convex inflationary potentials. The aim of our study is to compare the predictions of the theory of the origin of the universe from the quantum multiverse against the recent measurements of the Planck experiment in order to test this theory. In a companion paper we test the possibility that if the standard model of cosmology were extended to allow for new physics, such as the theory of the origin of the universe from a quantum multiverse discussed here, then ruling out these models of inflation may not be as straightforward in the extended picture. 

The paper is organized as follows: in the next Section we present the Starobinsky model as a representative of a concave downwards potential, and the modified inflaton field solution, the modified power spectrum, and the modified tensor-to-scalar ratio $r$, with all modifications derived from the quantum entanglement for our branch of the wavefunction of the universe with the others \cite{lmh}, in the context of the theory of the origin of the universe from the quantum multiverse. In Section \ref{method} we present the method of analyzing the predictions of this theory against data; we provide the results and the likelihood plots in Section \ref{results}, and we conclude in Section \ref{sec:conclusions}.

\section{The Modified Starobinsky Potential}
\label{sec:model}


The Starobinsky inflationary model \cite{staro} is an example of a concave downwards potential, where the field is rolling away from its minimum and $V'' < 0$. Therefore fluctuations can be unstable away from the flat part of the potential near the top, and since the field is away from its minimum, the saddle point expansion of the action on which the semiclassical approximation of quantum cosmology is based, is invalid. This class of potentials was studied in detail in \cite{lmh} for the theory of the origin of the universe from a quantum multiverse. Let us unwrap some notation and recap the main results of \cite{lmh} that we need in order to test them against data below.

In this theory, the wavefunction of the universe propagates through the landscape of vacua \cite{landscape}. The wavefunction operates initially on a two dimensional minisuperspace of 3-geometries denoted by $a$ and the landscape moduli $\phi$ which serves as a collective coordinate for the landscape with a disordered potential energy $V(\phi)$, which consists of a large number of vacua all with different energies. Solutions to the wavefunction are found by solving a Wheeler DeWitt (WdW) equation. Perturbing around the 3-geometries and the field vacua, results in an infinite number of fluctuations $f_n$. The long wavelength fluctuations are weakly coupled to the system, therefore they provide the ideal environment for triggering decoherence among the branches (solutions) of the wavefunction, in our case the system, and for inducing the quantum to classical transition of each branch. Including the backreaction of these fluctuations in the WdW equation results in a Master equation and an infinite sized midi superspaced now spanned by ${a, \phi, f_n }$. Details of the formalism and the solutions can be found in \cite{archillmh, richlmh, tomolmh, lmh}.

The distribution of landscape vacua is disordered. This system is quite similar mathematically to spin glass and quantum dots. The local parameter $b$ characterizes the depth, i.e. the energy of the particular vacuum site on which a branch has localized. This energy labeled by $b$ is presumably given by the local SUSY breaking scale of that vacuum site. Since vacua are at different energies, distributed in a disordered way, then the local parameter $b$ varies from vacua to vacua. In short, each landscape vacuum has two parameters, the global parameter (which is the same for all of the vacua) the string scale that we here take to be the Planck Mass $M_p$, and the local 'SUSY breaking' scale $b$. Due to disorder, the wavefunction branches undergo Anderson localization. They localize at different vacua with different energies. One of these branches becomes our universe as it decoheres from the other branches and undergoes a quantum to classical transition, thanks to its interaction with the environment made of the long wavelength fluctuation modes. But coherence and decoherence are closely related, and since we are studying a quantum system, then by unitarity, information about the quantum entanglement of our branch with others, is not wiped out and leaves its imprints in our classical sky after the universe has become classical. We have a coherent picture of how the branch evolves from the landscape and becomes a classical universe. We also have the quantum cosmology formalism that allows us to calculate these solutions and, to calculate the quantum entanglement of our branch with others. The latter enables us to derive how the observables in our sky predicted by inflation, get modified as a result of this additional source to perturbations and the gravitational potential of our universe, the quantum entanglement source. We consider here the Starobinsky model as an example of a concave downwards potential. We use in our analysis the modifications from quantum entanglement derived in \cite{lmh} when these potentials are embedded in the framework of a landscape multiverse.

To deal with the concave downwards potentials, the calculation in \cite{lmh} was performed in the 'Euclidean region' when implementing the steepest descent expansion and calculating corrections, which inverts the potential upside down as explained in detail in \cite{lmh}.

The highly nontrivial corrections to the concave downwards potentials with $V''<0$ like the Starobinsky model or hilltop potentials are such that $V_{eff} = V + \frac{1}{2} \frac{V^2}{9M^4} F[b,V]$ and $m^2 =Abs[V'']$ where the entanglement information is contained in the nonlocal term $F[b, V(\phi)]$ derived from quantum entanglement \cite{lmh}. Einstein equations get modified accordingly since the inflaton potential $V$ is now replaced by $V_{eff}$. The Friedmann equation for concave downwards potentials like the Starobinsky model \cite{staro}, with the notation $f(b, V) = \frac{V^2}{18 M_{p}^4} F[b,V]$, is

\begin{equation}
3 M_{p}^2 H^2 = V_{eff} = V +f[b,V].
\label{hubble}
\end{equation}

The field solution is also modified by replacing $V$ with $V_{eff}$ in the field equation. We have

\begin{equation}
3 H d\phi /dt = -\frac{\partial V_{eff}}{\partial \phi}.
\label{fieldsol}
\end{equation}

Using $d\phi/dt = H d\phi/dlogk$, the field equation \ref{fieldsol}, from Eq.\ref{hubble}, becomes

\begin{equation}
dN = \frac{V_{eff}}{M_{p}^2 V'_{eff}} d \phi = - d log k.
\label{fieldwithk}
\end{equation}

The correction term is small compared to the leading term $V$, so the slow roll conditions on the slow roll parameters $(\eta \epsilon)$ hold even with correction terms. Since the field is in the slow roll regime (i.e. it satisfies the condition of \cite{katieguth}, $\frac {V(\phi_{i}) - V(\phi_{f})}{\nabla \phi^{4}} \le 10^{-7}$ with $\phi_{i}, \phi_{f}$ the value of the field at the start and end of slow roll and $\nabla \phi$ the difference between the two), then although the correction terms are functions of $\phi$ we can approximate them with their value halfway through the slow roll, and the integral in Eq.~\ref{fieldwithk} as: $\int \frac{V_{eff}}{V'_{eff}} d\phi \simeq \left(\frac{1+ f/V}{1+ df/dV}\right)\int \frac{V}{V'} d\phi$.

Equation \ref{fieldwithk} gives us the field as a function of $k$, or equivalently as a function of the number of efolds $N$, therefore by requiring for example $N=60$ we can figure out the start of slow roll. That is, we can integrate $dN$ from the end $\phi_{end}$ to the start of slow roll $\phi_i$ to get the total number of efolds $N $.

\subsection{The Starobinsky Model}

The Starobinsky potential \cite{staro} can be reparameterized in terms of a scalar field $\phi$ to be given by a potential \cite{staro}

\begin{equation}
V(\phi) = V_0 \left(1-e^{-\sqrt{\frac{2}{3}}\frac{\phi}{M_P}}\right)^2
\label{starobinsky}
\end{equation}
and its derivative with respect to the field

\begin{equation}
V'(\phi) = 2\sqrt{\frac{2}{3}} \frac{V_0}{M_P} \left(1-e^{-\sqrt{\frac{2}{3}}\frac{\phi}{M_P}}\right)e^{-\sqrt{\frac{2}{3}}\frac{\phi}{M_P}}.
\end{equation}

The effective potential of the Starobinsky model modified by the quantum entanglement correction of the multiverse, is $V_{eff}(\phi)=V(\phi) + f(b, V[\phi])$

where $f[b,V]$ is given by:

\begin{equation}
f(\phi)=\frac{1}{2} \left[ \frac{V(\phi)}{3M_P^2} \right]^2 F(b, V[\phi]).
\end{equation}

Notice that the correction term in $V_{eff}$ has a $+$ sign in front as explained in \cite{lmh}. We will denote the reduced Planck mass by $M = 2.4 \times 10^{18}$ to distinguish it from the Planck mass $M_p$.
The nonlocal function $F[b,V]$ above, which depends on the global string scale, the inflaton potential $V(\phi)$ and the 'SUSY breaking' scale $b$ of the particular landscape vacuum where this branch sits is derived in \cite{lmh} to be

\begin{equation}
F(b,V[\phi]) = \frac{3}{2}\left(2+\frac{m^2M_P^2}{V(\phi)}\right)log\left(\frac{b^2M_P^2}{V(\phi)}\right)-\frac{1}{2}\left(1+\frac{m^2}{b^2}\right)e^{-\frac{3b^2M_P^2}{V(\phi)}}.
\end{equation}

The mass squared term in this expression is given by the magnitude of the curvature of the potential, and it is positive. Concave potentials have negative curvature and the calculation is performed in Euclidean section as mentioned above \cite{lmh}. When $F[b,V]$ is derived and we rotate back to Lorentz section then we have \cite{lmh} 
$m^2 = |V''(\phi_{in})|  >0$
which for the Starobinsky potential \ref{starobinsky} is

\begin{equation}
m^2 \simeq \frac{V_{0}}{M_{p}^2} Abs[ e^{- \sqrt{\frac{2}{3}}\frac{\phi}{M_p}}\left(1 - 2e^{- \sqrt{\frac{2}{3}} \frac{\phi}{M_p}} \right)] 
\label{staromass}
\end{equation}

$N$, the total number of efolds is obtained by integrating Eq. \ref{fieldwithk} from the start of slow roll $\phi_{in}$ to the end of slow roll $\phi_{end}$. So for $N \simeq 60$, we get $\phi_{in}\simeq 5.8 M_p$ and from Eq.\ref{staromass} we estimate $m^2 \simeq \frac{V_0}{M^{2}_{p} N}$.

The derivatives for the effective potential:

\begin{equation}
\frac{df}{dV}= \frac{V}{9 M_{p}^4} \left(F[b,V] + \frac{V}{2} dF/dV \right)
\end{equation}
where the expression for $F[b,V]$ is above, and $dF/dV$ is
\begin{equation}
\frac{dF}{dV} = -\frac{3\left(m^2 M_{p}^2 -M^2 (b^2 +m^2) Exp[-\frac{3b^2 M_{p}^2}{V}] -2V - m^2 M_{p}^2 log[\frac{b^2 M_{p}^2}{V}]\right)}{2 V^2}.
\label{dFdv}
\end{equation}

Putting this together we get $V'_{eff}$

\begin{equation}
V'_{eff}(\phi)=V'(\phi)\left(1+\frac{df}{dV}\right).
\end{equation}

Lets move on to field solutions. We can easily integrate Eq. \ref{fieldwithk} to get the field solution, which is what we did numerically for the analysis in this paper. However to gain some intuition in the field behaviour let us attempt an analytical solution. Note that during the slow roll, the potential is nearly a constant and therefore the correction term can be pulled out of the integral as explained below Eq. \ref{fieldwithk}.

\begin{equation}
\sqrt{\frac{2}{3}}\frac{1}{M_p}(\phi -\phi_{end})-\left(e^{\sqrt{\frac{2}{3}}\phi/M_{p}} - e^{\sqrt{\frac{2}{3}}\phi_{end}/M_{p}}  \right)= 4/3 log [k/k_{ref}]\frac{(1 + \frac{df(b, V[\phi(k)])}{dV[\phi(k)]})}{(1+ \frac{f[b, V[\phi(k)]}{V[\phi(k)]})}.
\label{noinvert}
\end{equation}

This is the exact solution. But it is a bit problematic because it is a transcendental function, thus not easy to invert. Note the correction term multiplying $log (k/k_{ref})$. Taking $f=0, df/dV=0$ recovers the unmodified field solution $\phi_0$ for the pure Starobinsky potential.

\begin{equation}
\sqrt{\frac{2}{3}}\frac{1}{M_p}(\phi_0 -\phi_{0 end})-\left(e^{\sqrt{\frac{2}{3}}\phi_{0}/M_{p}} - e^{\sqrt{\frac{2}{3}}\phi_{0 end}/M_{p}}  \right)= 4/3 log [k/k_{ref}]
\end{equation}
where $k_{ref}$ is a fiducial scale for the mode that we here take $k_{ref} = 0.002 hMpc^{-1}$.
Since we cannot invert Eq.~\ref{noinvert} analytically, then we need to somehow approximate it, or, alternatively, invert it numerically. To get an idea, let us give below the analytic approximation of the above exact solution, with notations $\phi_{end}$ and  $\phi_{in}= log[4/3 (N - log[k/k_{ref}])]$ for the end and start values of the field during slow roll.

The unmodified field solution is approximately

\begin{equation}
\phi_{0Analytic}(k)= \sqrt{\frac{3}{2}}M log[\frac{4}{3}\left(N - log[k/k_{ref}]\right) + Exp[\phi_{end}] + \phi_{in}(k) - \phi_{end}]
\end{equation}
and the approximate modified field solution
\begin{equation}
\phi_{Analytic}(k)= \sqrt{\frac{3}{2}}M log[\frac{4}{3}\left(N - log[k/k_{ref}] \right)\frac{(1+df(\phi_{0Analytic})/dV)}{1 + f(\phi_{0Analytic})/V} + Exp[\phi_{end}] + \phi_{in}(k) - \phi_{end} ].
\label{modifiedfield}
\end{equation}

Note that the correction term for the field goes as $(1+ df/dV)/(1+ f/V)$. 

All unmodified quantities like $r_0[k]$, the scalar spectral index $n_0[k]$, the power spectrum $P_{0}[k]$ etc., are evaluated with respect to the unmodified field $\phi_0$. All modified quantities such as $r[k], n[k]$ and the power spectrum $P[k]$, are evaluated with respect to the modified field above $\phi$.

Using the modified field solution to calculate the spetrum, we have

\begin{equation}
P_\zeta(k)=\frac{1}{24\pi^2M_P^6}\left[ \frac{V_{eff}(\phi)^3}{V'_{eff}(\phi)^2} \right] = \frac{1}{24\pi^2M_P^6}\left[ \frac{V(\phi)^3}{V'(\phi)^2} \right]\frac{(1+f/V)^3}{(1 +df/dV)^2} .
\end{equation}

It can be seen that the modified power spectrum is related to the unmodified one , $P_0[k]$, approximately by 
\begin{equation}
P[\phi(k)] \approx P_{0}[\phi_{0}(k)] \frac{(1+ f/V)^3}{(1+ df/dV)^2}.
\end{equation}

We should bear in mind that there is a difference between the modified and the unmodified spectra, because the former is evaluated from the modified field solution $\phi(k)$, while the latter is evaluated from the unmodified field solution $\phi_{0}(k)$ where $\phi(k) \ne \phi_{0}(k)$. Therefore the above expression relating $P[\phi(k)]$ to $P_{0}[\phi_{0}(k)]$ is only approximate with this caveat. Also recall that the reduced Planck mass $M$ used for these expressions, is obtained from the original one $M_p$ by dividing it with $1/ \sqrt{8 \pi}$.

Finally let us estimate the tensor to scalar ratio $r[k]$
\begin{equation}
r[k]= 8 M_{p}^2 \left(\frac{V'_{eff}(\phi)}{V_{eff}(\phi)}\right)^2 =  8 M_{p}^2 \left(\frac{V'(\phi)}{V(\phi)}\right)^2 (\frac{( 1 + df/dV )}{( 1 + f/V )})^2
\end{equation}
and the unmodifed one by
\begin{equation} 
r_{0} [k] = 8 M_{p}^2 \left(\frac{V'(\phi_0)}{V(\phi_0)}\right)^2 .
\end{equation}

The tensor to scalar ratio $r[k]$ calculated above can be estimated from $V'$ and $V'_{eff}$, in a straightforward manner. Again, bearing in mind the caveat that the modified field solution enters the modified tensor index $r$,  while $\phi_0$ enters $r_{0}[k]$ estimations, then we have an approximate relation, 
$r[k] \approx r_{0}[k]  (\frac{( 1 + df/dV )}{( 1 + f/V )})^2$.

From here we calculate $n[k] - 1 = dlog(P_{zeta}[k])/dlog(k)$ from the expression for $P[k]$ above. Then the unmodified scalar tensor is $n_{0}[k] -1 =  dlog(P_0[k])/dlog(k)$ where recall that the $_0$ notation always denotes the unmodified field, power spectrum, and scalar or tensor spectral indices.

\section{Analysis Method} \label{method}
We explore the modified Starobinsky model by considering as a baseline the cosmological parameters listed next. These are the $4$ standard parameters of the $\Lambda$CDM and $3$ inflationary parameters: the baryon $\Omega_bh^2$ and the cold dark matter energy densities $\Omega_ch^2$; the reionization optical depth $\tau$; the ratio between the sound horizon and the angular diameter distance at decoupling $\Theta_{s}$; the natural logarithm of the SUSY-breaking scale associated with the landscape effects $log(\sqrt{8\pi}b[GeV])$ (note the scaling factor $\sqrt{8 \pi}$ coming from the reduced Planck mass which rescales $b$ since this is defined in terms of the normal Planck mass in the correction term); the energy scale of the inflation $10^{10}V_0/M^4$, and the contribution of the primordial gravitational waves with a tensor-to-scalar ratio of amplitude $r$ at the pivot scale $k_{ref}=0.002 hMpc^{-1}$. 

\begin{figure}
\centering
\includegraphics[width=10cm]{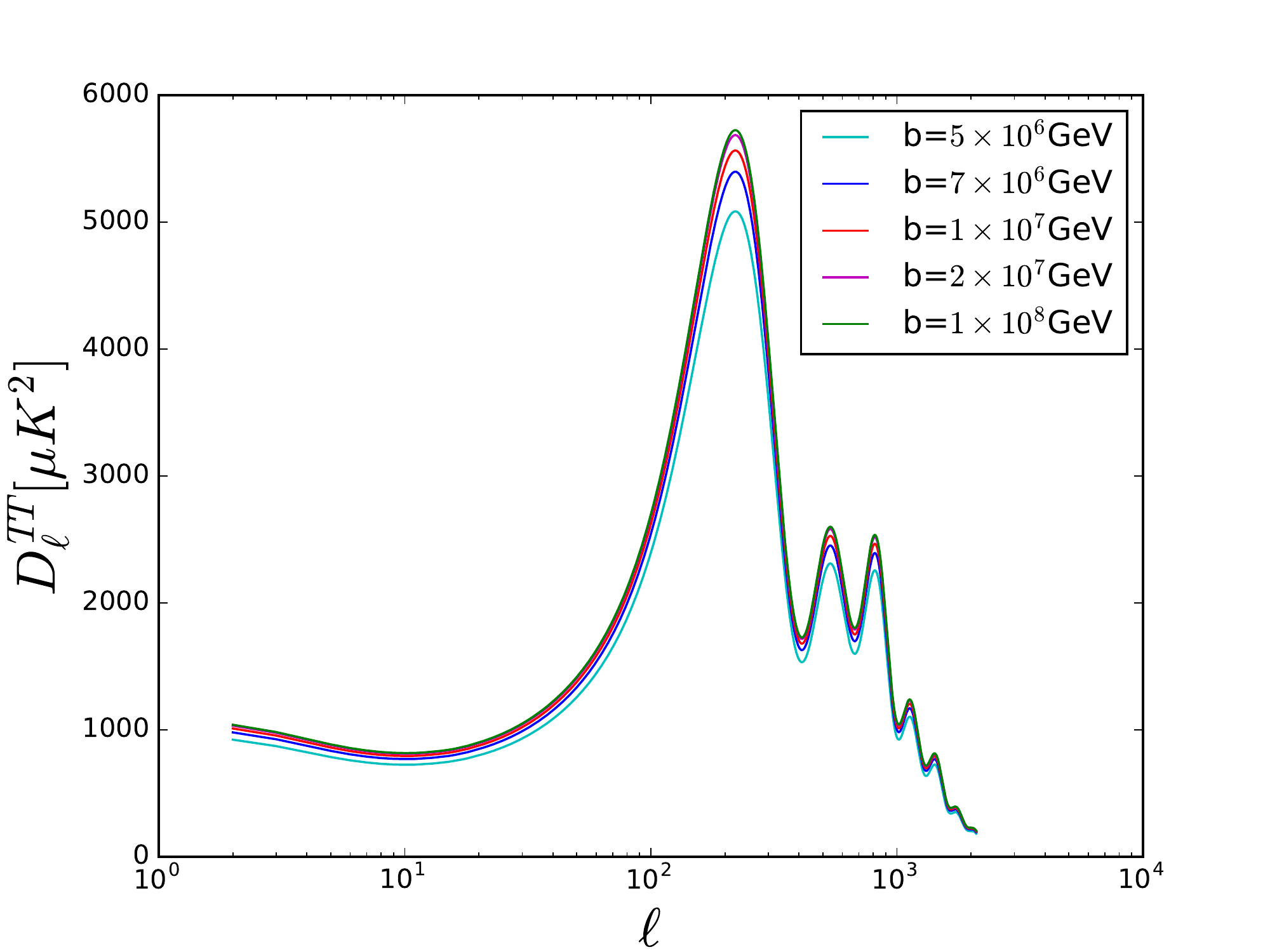}
\includegraphics[width=10cm]{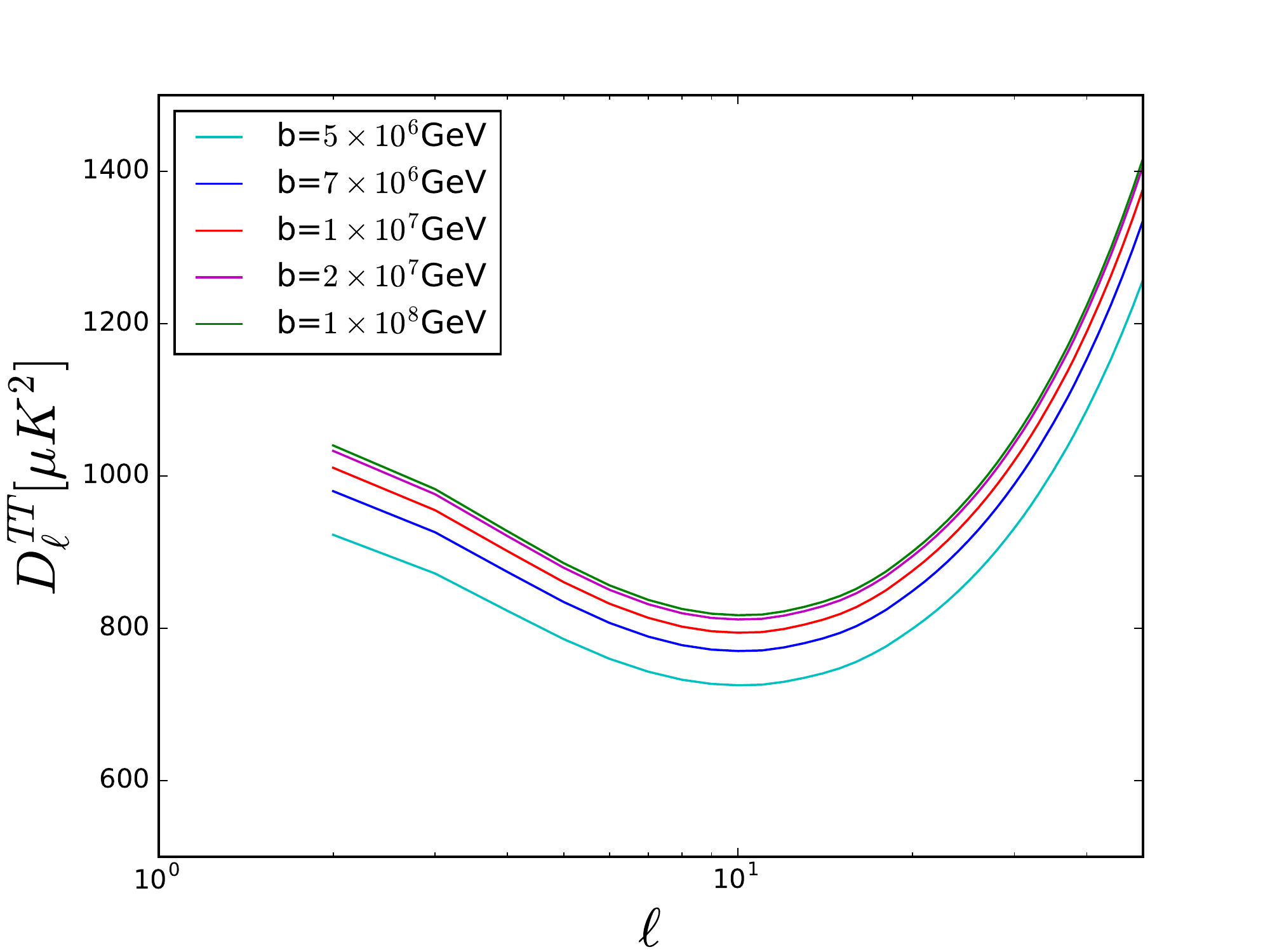}
\caption{The temperature CMB angular power spectrum by varying the SUSY-breaking scale $b$ associated with the landscape effects, for the modified Starobinsky model.}
\label{clparamST}
\end{figure}

\begin{figure}
\centering
\includegraphics[width=10cm]{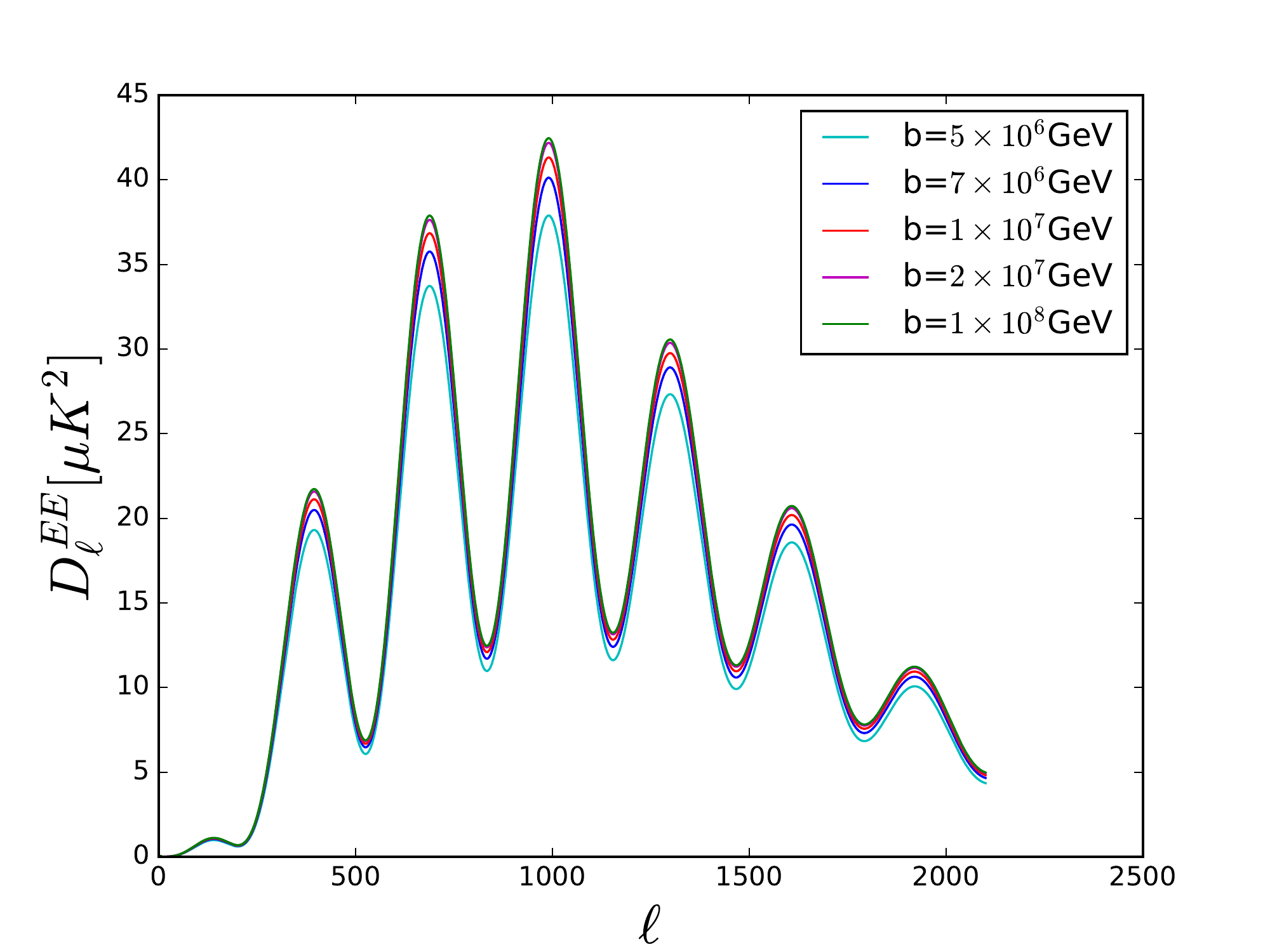}
\includegraphics[width=10cm]{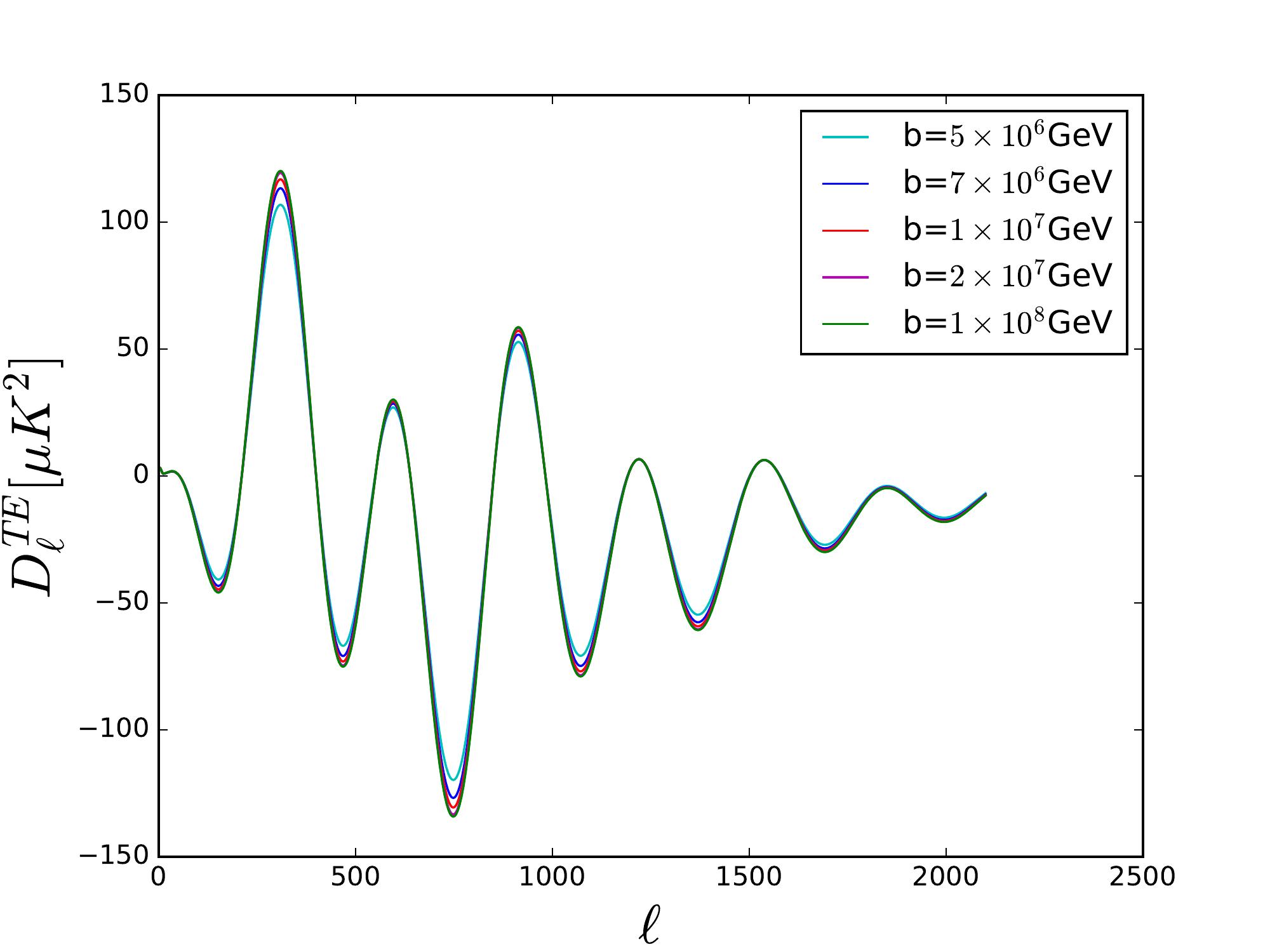}
\caption{The polarization CMB angular power spectra by varying the SUSY-breaking scale $b$ associated with the landscape effects, for the modified Starobinsky model.}
\label{clparamSP}
\end{figure}

\begin{figure}
\centering
\includegraphics[width=10cm]{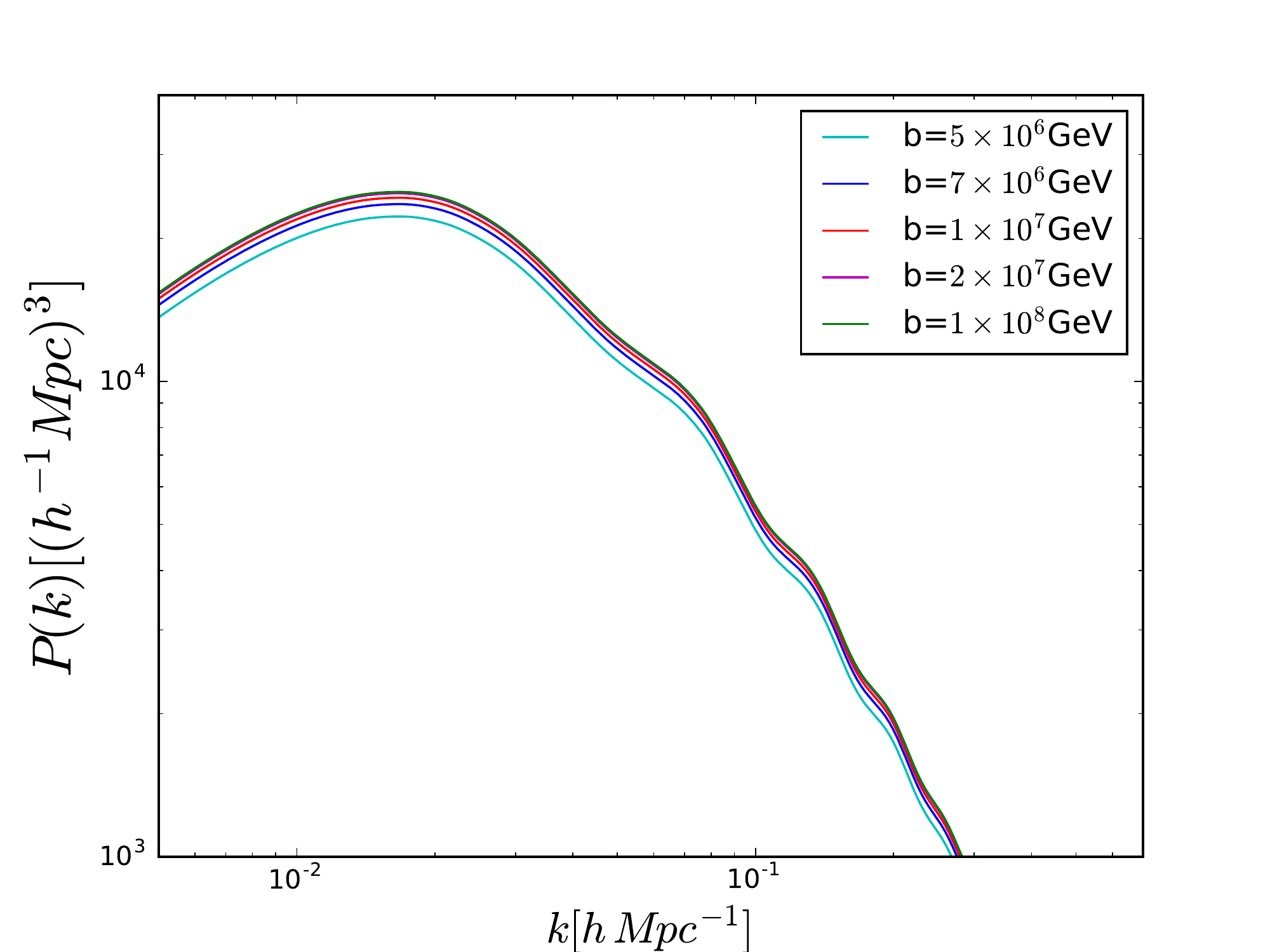}
\caption{The matter power spectrum by varying the SUSY-breaking scale $b$ associated with the landscape effects, for the modified Starobinsky model.}
\label{clparamSM}
\end{figure}

Moreover, we also consider extensions to these baseline models, by adding one more parameter at a time, such as: the effective number of relativistic degrees of freedom $N_{\rm eff}$ and the dark energy equation of state $w$. Finally, we vary both extension parameters at the same time. 
All the parameters we consider in our analysis are explored within the range of the conservative flat priors reported in Table~\ref{priors}. Figs.~\ref{clparamST},~\ref{clparamSP} and ~\ref{clparamSM} show the manner in which the SUSY-breaking scale $b$ associated with the landscape effects, affects the CMB temperature, polarization and matter power spectra, respectively.

\begin{table}
\begin{center}
\begin{tabular}{|c|c|}
\hline
Parameter                    & Starobisky \\
\hline
$\Omega_{\rm b} h^2$         & $[0.005,0.1]$ \\
$\Omega_{\rm cdm} h^2$       & $[0.001,0.99]$ \\
$\Theta_{\rm s}$             & $[0.5,10]$ \\
$\tau$                       & $[0.01,0.8]$ \\
$log(\sqrt{8\pi}b[GeV])$                        & $[16, 23]$ \\
$10^{10}V_0/M^4$         & $[1.7,2.4]$ \\
$r$ & $[0,3]$ \\
$N_{\rm eff}$ & $[0.05,10]$ \\
$w$ & $[-3.0,0.3]$ \\
\hline
\end{tabular}
\end{center}
\caption{External priors on the cosmological parameters assumed in this work.}
\label{priors}
\end{table}


We find the constraints on these parameters below by considering several recent cosmological probes.
Firstly, we analyze the ``Planck TT + lowTEB'' data, i.e. the full range of the 2015 temperature power spectrum ($2\leq\ell\leq2500$) combined with the low-$\ell$ polarization power spectra in the multipoles range $2\leq\ell\leq29$ provided by the Planck collaboration \cite{Aghanim:2015xee}.
Secondly, we include the high multipoles Planck polarization data \cite{Aghanim:2015xee}, in the range $30\leq\ell\leq2500$, that we will call ``Planck TTTEEE + lowTEB''. 
Then, we replace the lowTEB data with a gaussian prior on the reionization optical depth $\tau=0.055\pm0.009$, as obtained recently from Planck HFI measurements \cite{newtau}, and we will call this prior ``tau055''.
In addition, we consider the 2015 Planck CMB lensing reconstruction power spectrum $C^{\phi\phi}_\ell$ \cite{Ade:2015zua}, and we will refer to this dataset as ``lensing''.
We add the baryonic acoustic oscillation data from 6dFGS \cite{beutler2011}, SDSS-MGS \cite{ross2014}, BOSSLOWZ \cite{anderson2014} and CMASS-DR11 \cite{anderson2014} surveys as was done in \cite{planckparams2015}, and we will call this dataset "BAO".
Furthermore, we consider the ``BKP'' dataset, i.e. the CMB polarization $B$ modes constraints provided by the 2014 common analysis of Planck, BICEP2 and Keck Array \cite{BKP}.
Finally, we will include a gaussian prior on the Hubble constant $H_0=73.2\pm1.7$ km/s/Mpc, quoting the directly measured value provided recently by Riess et al. \cite{R16}, and we will refer to this prior as "H073p2".

In order to analyze statistically these data exploring the modified Starobinsky model for the entanglement, we have used the June 2016 version of the publicly available Monte-Carlo Markov Chain package \texttt{cosmomc} \cite{Lewis:2002ah}, with a convergence diagnostic based on the Gelman and Rubin statistic. In this version, we modified the CAMB code \cite{Lewis:1999bs}, to include the primordial power spectrum of our model. It implements an efficient sampling of the posterior distribution using the fast/slow parameter decorrelations \cite{Lewis:2013hha}, and it includes the support for the Planck data release 2015 Likelihood Code \cite{Aghanim:2015xee} (see \url{http://cosmologist.info/cosmomc/}). 

\section{Results}\label{results}
\subsection{Data analysis}\label{data}

\begin{table}[!]
\begin{center}\footnotesize
\scalebox{0.78}{\begin{tabular}{lcccccc}
\hline \hline
      Planck TT    & & && &&\\                     
         & + lowTEB     &        + lowTEB + BAO  & + lowTEB + lensing      &  + lowTEB + BKP&+tau055 \\  
\hline
\hspace{1mm}\\

$\Omega_{\textrm{b}}h^2$& $0.02225\,\pm 0.00019 $& $0.02227\,\pm0.00019$    & $0.02223\,\pm0.00019 $& $0.02224\,\pm 0.00019$& $0.02222\,\pm 0.00019$   \\
\hspace{1mm}\\

$\Omega_{\textrm{c}}h^2$& $0.1195\, \pm0.0012$& $0.11923\,\pm 0.00097$    & $0.1190\,\pm0.0011 $& $0.1196\,\pm0.0012$ & $0.1196\,\pm0.0012$   \\
\hspace{1mm}\\

$\tau$& $0.078\,^{+0.018}_{-0.016}$& $0.079\,\pm0.016$    & $0.063\,\pm 0.012$& $0.081\,\pm 0.016$& $0.0619\,\pm 0.0086$    \\
\hspace{1mm}\\

$10^{10}V_0/M^4$& $2.209\pm0.075$& $2.209\pm0.072$    & $2.139\,\pm 0.049$& $2.220\,\pm 0.073$& $2.139\,^{+0.039}_{-0.047}$   \\
\hspace{1mm}\\

$log(\sqrt{8\pi}b[GeV])$& $>19.4$& $>19.5$    & $20.4\,\pm 1.5$& $>19.5$& $>19.4$    \\
\hspace{1mm}\\

$r$ &  $0.0029674\,_{-0.0000001}^{+0.0000024}$ &  $0.0029678\,_{-0.0000001}^{+0.0000016}$  & $0.0029677\,_{-0.0000001}^{+0.0000018}$&  $0.0029677\,_{-0.00000004}^{+0.0000018}$ &  $0.0029673\,_{-0.0000001}^{+0.0000026}$ \\
\hspace{1mm}\\

$H_0$ &      $67.41\pm0.53$&      $ 67.55\pm0.42$ & $ 67.62\,\pm0.50$   &  $ 67.37\,\pm 0.51$ &  $ 67.37\,\pm 0.53$  \\
\hspace{1mm}\\

$\sigma_8$   & $ 0.829\,\pm0.015$   & $ 0.828\,\pm0.014$   & $ 0.8144\,\pm 0.0084$ &  $ 0.832\,\pm0.014$ &  $ 0.8160\,\pm0.0089$ \\
\hspace{1mm}\\

\hspace{1mm}\\
\hline
\hline
      Planck TTTEEE    & & && &&\\                     
          & + lowTEB     &        + lowTEB + BAO  & + lowTEB + lensing      &  + lowTEB + BKP&+tau055 \\  
\hline
\hspace{1mm}\\

$\Omega_{\textrm{b}}h^2$& $0.02226\,\pm 0.00013 $& $0.02228\,\pm0.00013$    & $0.02226\,\pm0.00013 $& $0.02226\,\pm 0.00013$& $0.02224\,\pm 0.00014$  \\
\hspace{1mm}\\

$\Omega_{\textrm{c}}h^2$& $0.11954\, \pm0.00093$& $0.11928\,\pm 0.00079$    & $0.11912\,\pm0.00093 $& $0.11957\,\pm0.00094$ & $0.11965\,\pm0.00095$  \\
\hspace{1mm}\\

$\tau$& $0.081\,\pm0.015$& $0.082\,\pm0.015$    & $0.064\,\pm 0.011$& $0.083\,\pm 0.015$& $0.0629\,\pm 0.0086$     \\
\hspace{1mm}\\

$10^{10}V_0/M^4$& $2.224\pm0.068$& $2.223\pm0.067$    & $2.142\,^{+0.044}_{-0.054} $& $2.232\,\pm 0.068$& $2.147\,^{+0.041}_{-0.047}$   \\
\hspace{1mm}\\

$log(\sqrt{8\pi}b[GeV])$& $20.4\,\pm1.5$& $>19.6$    & $20.3\,\pm1.5$& $20.3\,^{+1.5}_{-2.0}$& $>19.3$  \\
\hspace{1mm}\\

$r$ &  $0.0029676\,_{-0.0000001}^{+0.0000021}$ &  $0.0029681\,_{-0.00000002}^{+0.0000011}$  & $0.0029677\,_{-0.0000001}^{+0.0000018}$&  $0.0029676\,_{-0.0000001}^{+0.0000020}$ &  $0.0029670\,_{-0.0000001}^{+0.0000032}$ \\
\hspace{1mm}\\

$H_0$ &      $67.39\pm0.41$&      $ 67.51\pm0.35$ & $ 67.56\,\pm0.42$   &  $ 67.38\,\pm 0.42$ &  $ 67.33\,\pm 0.43$  \\
\hspace{1mm}\\

$\sigma_8$   & $ 0.832\,\pm0.013$   & $ 0.831\,\pm0.013$   & $ 0.8152\,\pm 0.0081$ &  $ 0.834\,\pm0.013$ &  $ 0.8173\,\pm0.0082$   \\
\hspace{1mm}\\

\hspace{1mm}\\
\hline
\hline

\end{tabular}}
\caption{$68 \% $ c.l. constraints on cosmological parameters in our baseline $\Lambda$CDM+r scenario from different combinations of datasets with a modified Starobisky inflation.}
\label{table1}
\end{center}
\end{table}

\begin{figure}
\centering
\includegraphics[scale=1.0]{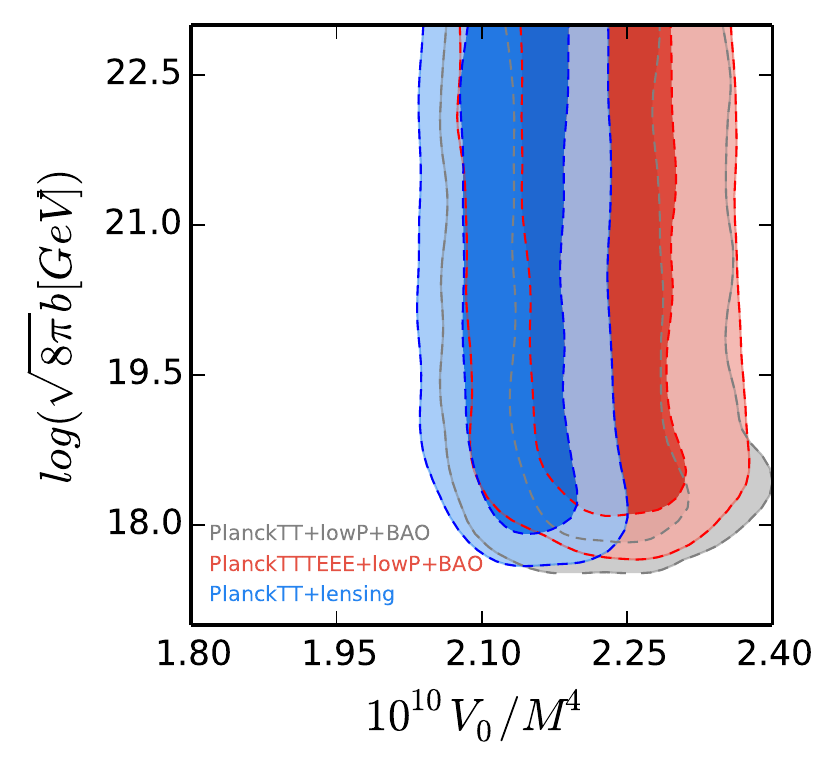}
\caption{Constraints at $68 \%$ and  $95 \%$ confidence levels on the $10^{10}V_0/M^4$ vs $log(\sqrt{8\pi}b[GeV])$ plane, in our modified $\Lambda$CDM+r Starobisky scenario.}
\label{figv0b}
\end{figure}

The result of all the explorations are given in Tables~\ref{table1}, \ref{table2}, \ref{table3} and \ref{table4}, where we report the constraints at $68 \% $ c.l. on the cosmological parameters. All the bounds that we will quote hereinafter there will be at $68\%$ c.l., unless otherwise expressed. These Tables differ for the cosmological scenario explored, respectively the $\Lambda$CDM+r, $\Lambda$CDM+r+$N_{\rm eff}$, $w$CDM+r and $w$CDM+r+$N_{\rm eff}$. 


\begin{figure}
\centering
\includegraphics[width=10cm]{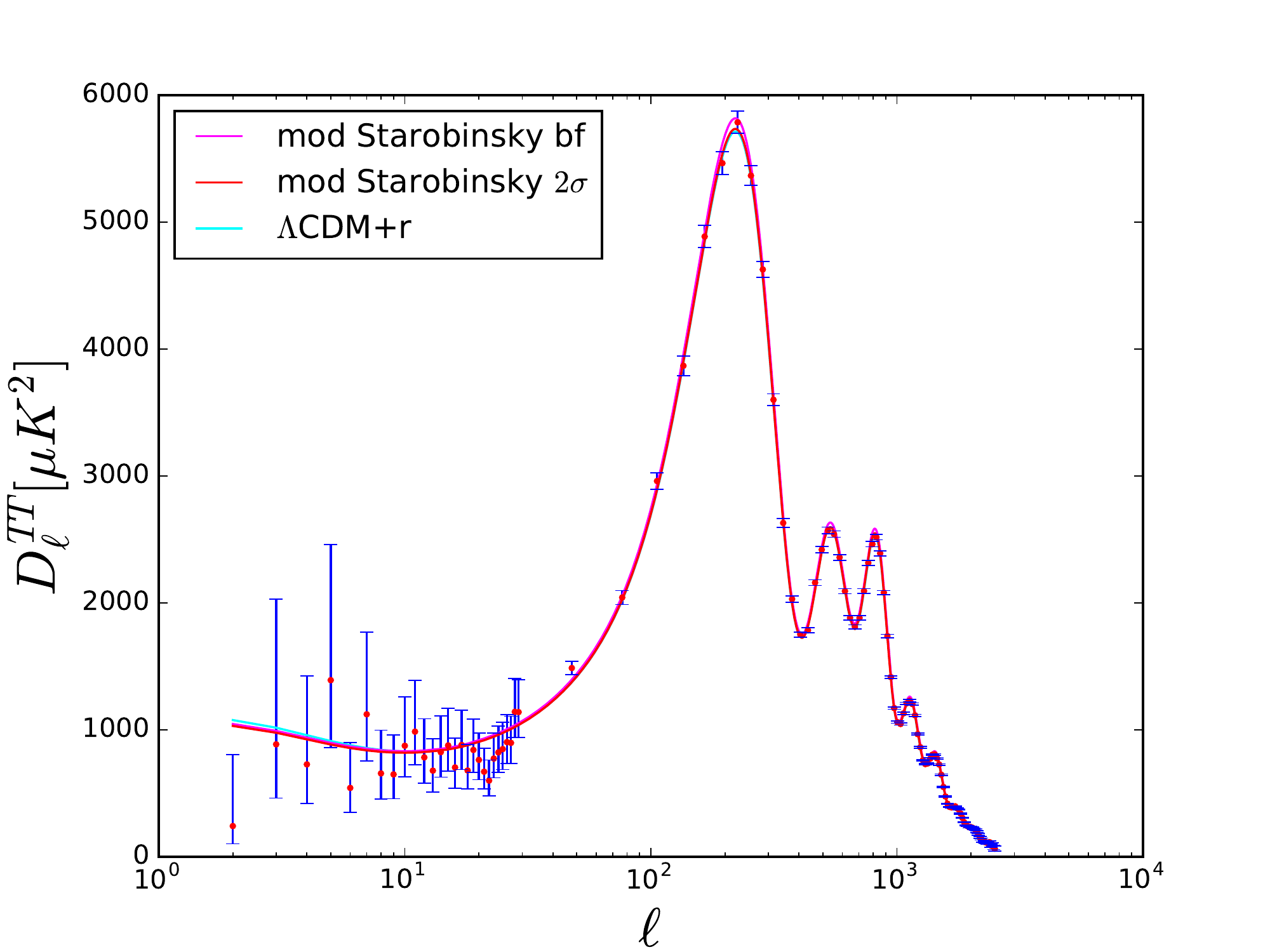}
\includegraphics[width=10cm]{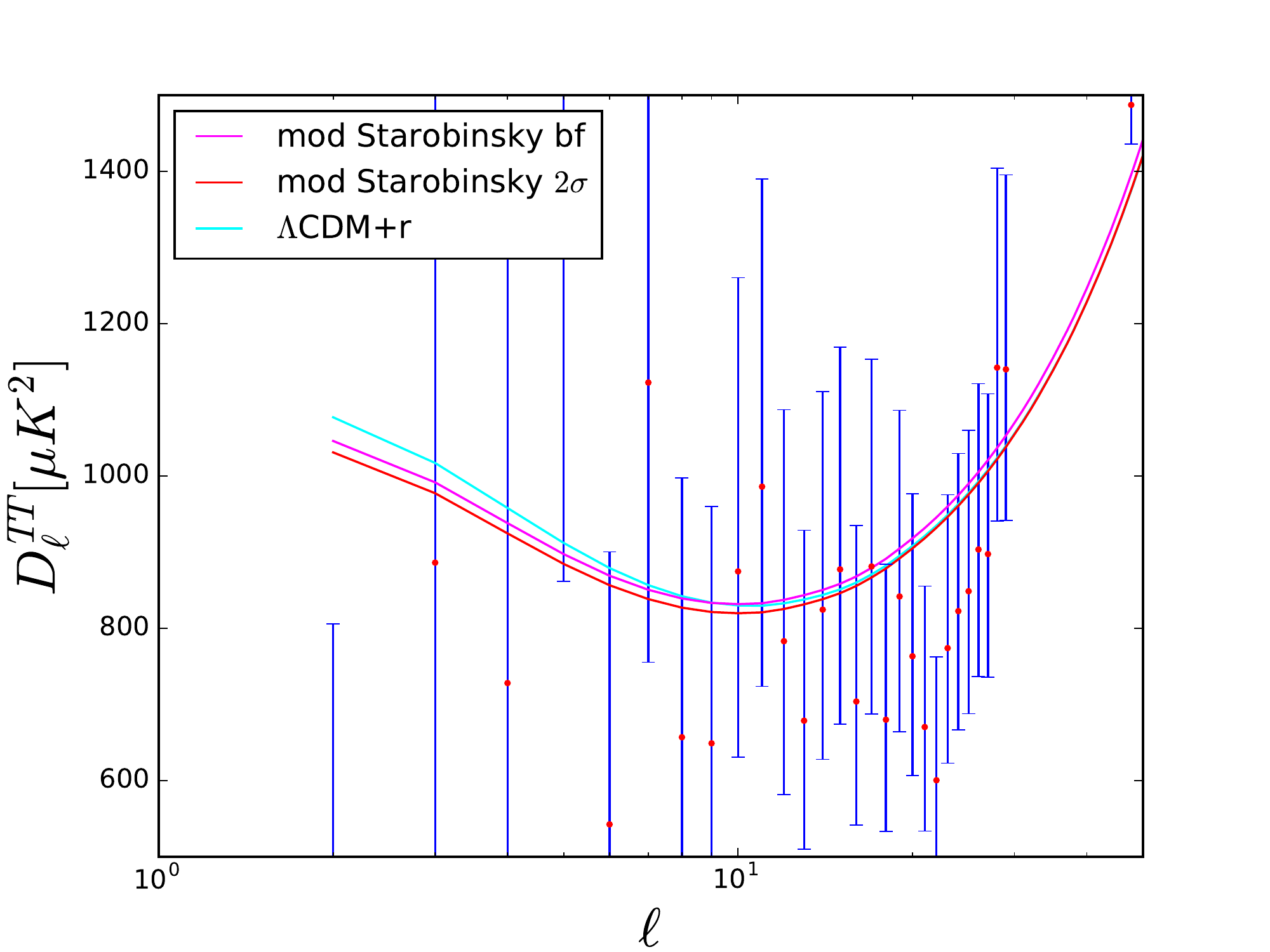}
\caption{Comparison of the temperature CMB angular power spectrum computed for the best-fit of our modified Starobinsky model (magenta), the lower $2\sigma$ limit of $b$ for our modified Starobinsky model (red), and the best-fit obtained with a minimal standard cosmological model $\Lambda$CDM+r (cyan), with Planck 2015 TT+lowTEB data (points with error bars). The main differences between the two models are at lower-$\ell$ and on the amplitude of the peaks, that the Starobinsky model, modified for the entanglement, prefer slightly higher and more in agreement with the data.}
\label{clbf}
\end{figure}

\begin{figure}
\centering
\includegraphics[width=10cm]{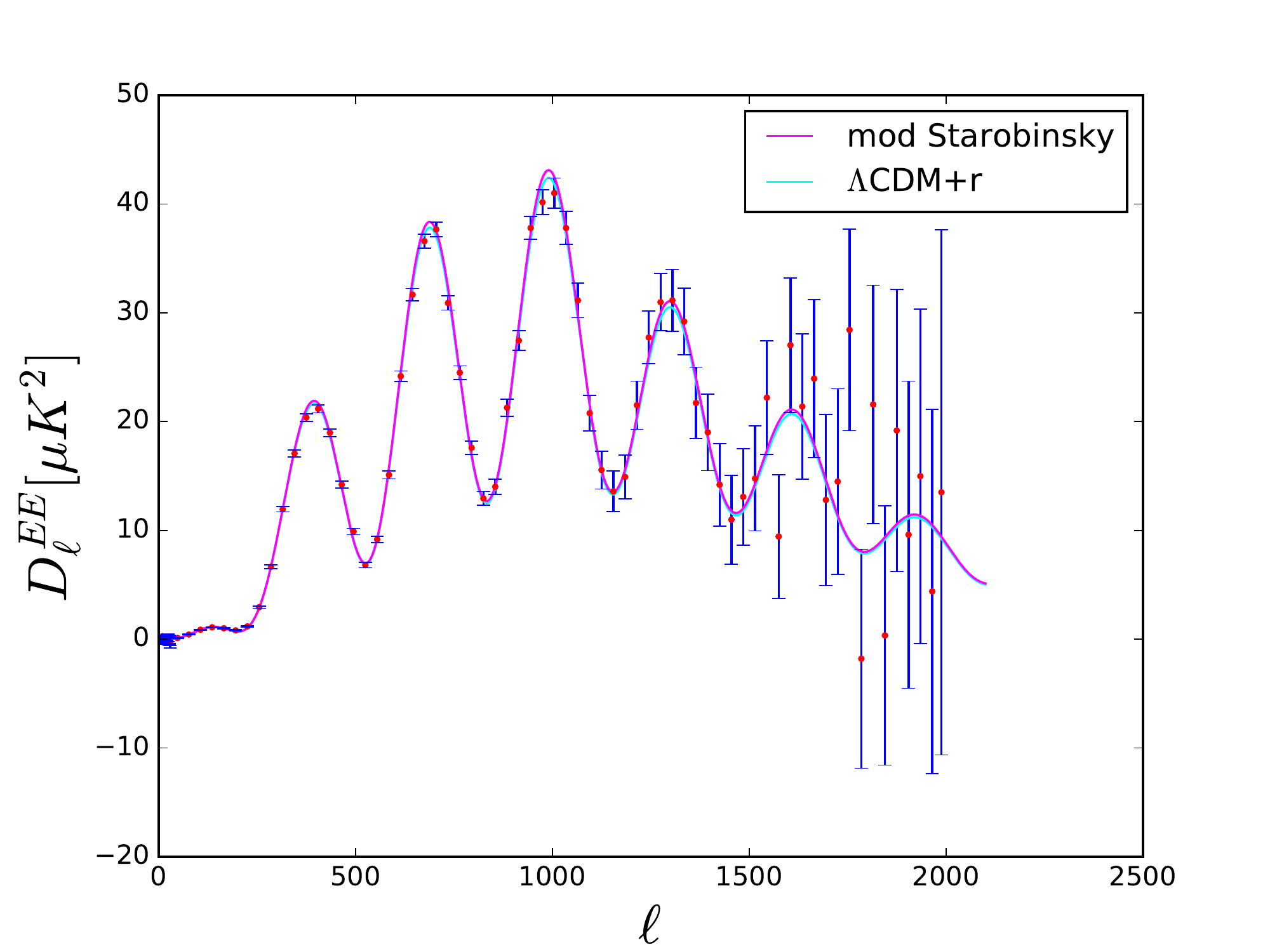}
\includegraphics[width=10cm]{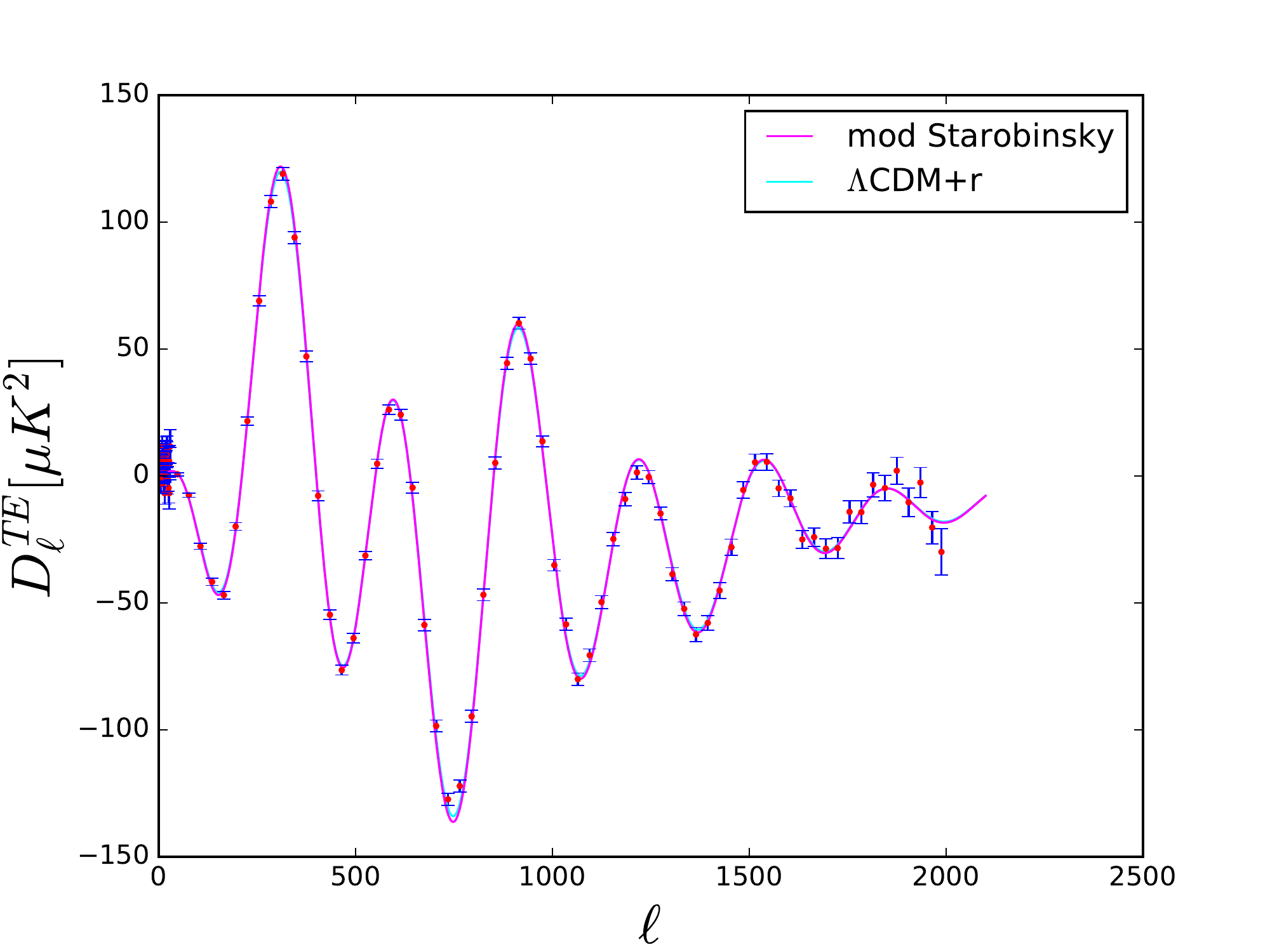}
\caption{Comparison of the polarization CMB angular power spectra computed for the best-fit of our modified Starobinsky model (magenta) and the best-fit obtained with a minimal standard cosmological model $\Lambda$CDM+r (cyan), with Planck 2015 TT+lowTEB data (points with error bars).}
\label{clbf_pol}
\end{figure}

Regarding the results of the Table~\ref{table1}, we notice that this modified version of the Starobinsky model produces a slightly better fit of the data, especially at the first peak and in the low-$\ell$ range, gaining about a $\Delta \bar \chi^2=2.5$, when considering Planck TT+lowTEB data and its combinations, and an improvement of about a $\Delta \bar \chi^2=3.2$, when considering Planck TTTEEE+lowTEB data and its combination, with respect to the minimal standard cosmological model $\Lambda$CDM+r (see Figs.~\ref{clbf} and ~\ref{clbf_pol}). In any case, in the modified Starobinsky model, we have one less degree of freedom, because $r$ and $n_S$ are derived parameters, and we are vary only $6$ parameters instead of $7$, as in the standard inflation. In particular, we have a better fit of the low multipole range, improving the $\Delta \bar \chi^2$ of about $1.4-1.7$ and improving the high multipole range by about $0.7-1.0$, depending on the considered combination of datasets. The high-$\ell$ improvement is probably due to the higher first peak preferred by the modified Starobinsky model for the entanglement, with respect to the minimal standard cosmological model $\Lambda$CDM+r.

When including the high-$\ell$'s polarization data, we have a further improvement of the $\bar \chi^2$ of the priors that we consider for the foregrounds, by imposing the modified Starobinsky scenario. In fact, we have a shift at more than one standard deviation towards higher values of the amplitude of the galactic dust for TE, in particular for the $217GHz$ and the $143\times217GHz$, that are in better agreement with the expected values. We find that $A^{dust,TE}_{143\times217}$ shifts from $0.34\pm0.08$ to $0.54\pm0.13$ and $A^{dust,TE}_{217}$ from $1.67\pm0.26$ to $1.91\pm0.28$ for Planck TTTEEE+lowTEB datasets.

By introducing this modified Starobinsky inflation, as can be seen from the comparison of Table~\ref{table1} and Table~\ref{tablelcdm}, we have very robust constraints for all the cosmological parameters with no significant departure from their values with respect to the standard case. Moreover, both $\Omega_ch^2$ and the Hubble constant $H_0$ have stronger constraints in the modified Starobinsky model than the pure inflation scenario, reducing the error bars by a half for all the dataset combination considered here. For example, we have that $\Omega_ch^2=0.1195\pm0.0022$ for Planck TT+lowTEB in the minimal standard cosmological model $\Lambda$CDM+r, which becomes $\Omega_ch^2=0.1195\pm0.0012$ in the modified Starobinsky model for the same combination of data, and where $H_0=67.42\pm0.99$ Km/s/Mpc becomes $H_0=67.41\pm0.53$ Km/s/Mpc. For this reason, the tension existing between the Planck data and the direct measurements of the Hubble constant \cite{R16} increases up to $3.2\sigma$. Moreover, if we look at the value of $S_8$, the other famous tension between the Planck data and the weak lensing experiments, as for example CFHTLenS \cite{Heymans:2012gg} and KiDS-450 \cite{Hildebrandt:2016iqg}, we find that this is still present. In fact, if we consider $S_8 \equiv \sigma_8 \sqrt{\Omega_m / 0.3}$, we have from the KiDS-450 collaboration $S_8=0.745 \pm 0.039$, and we find $S_8=0.849\pm0.020$, i.e. a $2.4\sigma$ tension.

When adding the lensing dataset, the optical depth shifts towards a lower value reducing its error bar more than one third. This parameter, which is equal to $\tau=0.067\pm0.017$ for Planck TTTEEE+lowTEB+lensing data in the minimal standard cosmological model $\Lambda$CDM, becomes, $\tau=0.064\pm0.011$ for the same combination of measurements in the modified Starobinsky model by quantum entanglement, in perfect agreement and with comparable error bars to the new value provided by the HFI data of the Planck collaboration.

Regarding the inflationary parameters that describe the theory analyzed here, we have a very precise and robust expectation for the derived value of the tensor-to-scalar ratio $r$, which is different from zero at several standard deviations. We find for this model, given the combination of obtained parameters, that $r\sim0.002968$. If we look at Fig.~\ref{figv0b} that shows the constraints at $68 \%$ and  $95 \%$ confidence levels on the $10^{10}V_0/M^4$ vs $log(b)$ plane, we can see that there exists a lower limit for $b$ at $b>1.2\times10^7 GeV$ at $95 \%$ c.l. and $V_0=(2.208\pm0.075)\times 10^{10} M_P^4$ at $68\%$ c.l. for Planck TT+lowTEB. An upper limit on $b$ was obtained in \cite{tomolmh} by considerations of the inflaton potential remaining nearly flat $\frac{\nabla V}{\nabla \phi^{4}} \le 10^{-7}$ even in the presence of quantum entanglement corrections given by $f[b, V]$. Interestingly, for some combination of datasets, as for example Planck TT+lowTEB+lensing datasets, some indication at $1\sigma$ of an upper and lower bound appear for the 'SUSY breaking' scale $b$,  $1.2\times10^7GeV<b<6.5\times10^8GeV$. We remember that from Fig.~\ref{figv0b} we can derive the 2D lower limit on $b$, that is $b>1.3\times10^7 GeV$ at $1\sigma$ level. However, by marginalizing this result over all the cosmological parameters, i.e. also over $V_0$ to obtain the posterior distribution of $b$, we find $b>5.7\times10^7 GeV$ at $1\sigma$ level, as it is reported in Table~\ref{table1}. The best fit of the parameter $b$ will be, instead, the value for which we have the minimum value for the $\chi^2$, and this is different from the mean value, if the parameter is not totally gaussian distributed.

\begin{table}[!]
\begin{center}\footnotesize
\scalebox{0.8}{\begin{tabular}{lccccc}
\hline \hline
      Planck TT    & &  &&\\                     
         & + lowTEB     &        + lowTEB + BAO  & + lowTEB + lensing  &+ lowTEB + H073p2\\  
\hline
\hspace{1mm}\\

$\Omega_{\textrm{b}}h^2$& $0.02224\,\pm 0.00019 $& $0.02226\,\pm0.00018$    & $0.02222\,\pm0.00019 $ & $0.02232\,\pm 0.00018$   \\
\hspace{1mm}\\

$\Omega_{\textrm{c}}h^2$& $0.1201\, \pm0.0033$& $0.1191\,\pm 0.0028$    & $0.1185\,\pm0.0030 $ & $0.1206\,\pm 0.0033$   \\
\hspace{1mm}\\

$\tau$& $0.078\,\pm0.017$& $0.079\,\pm0.017$    & $0.064\,\pm 0.013$& $0.077\,\pm 0.016$    \\
\hspace{1mm}\\

$10^{10}V_0/M^4$& $2.209\pm0.072$& $2.210\pm0.073$    & $2.136\,\pm 0.050 $&  $2.208\,\pm0.072$   \\
\hspace{1mm}\\

$log(\sqrt{8\pi}b[GeV])$& $20.5\,\pm1.5$& $20.4\,\pm1.5$    & $>19.7$& $>19.5$    \\
\hspace{1mm}\\

$r$ &  $0.0029680\,_{-0.00000003}^{+0.0000013}$ &  $0.0029679\,_{-0.00000002}^{+0.0000015}$  & $0.0029682\,_{-0.00000002}^{+0.0000010}$&   $0.0029680\,_{-0.00000003}^{+0.0000013}$   \\
\hspace{1mm}\\

$N_{\rm eff}$ &  $3.06\pm0.13$&      $ 3.04\pm0.12$ & $ 3.02\,\pm0.12$   &    $ 3.12\,\pm 0.12$ \\
\hspace{1mm}\\

$H_0$ &      $67.43\pm0.58$&      $ 67.53\pm0.55$ & $ 67.53\,\pm0.58$    &   $ 68.00\,\pm 0.56$ \\
\hspace{1mm}\\

$\sigma_8$   & $ 0.831\,\pm0.015$   & $ 0.828\,\pm0.014$   & $ 0.8134\,\pm 0.0092$  &    $ 0.830\,\pm0.015$  \\
\hspace{1mm}\\

\hspace{1mm}\\
\hline
\hline
      Planck TTTEEE    & &  &&\\                     
          & + lowTEB     &        + lowTEB + BAO  & + lowTEB + lensing    &+ lowTEB + H073p2\\  
\hline
\hspace{1mm}\\

$\Omega_{\textrm{b}}h^2$& $0.02227\,\pm 0.00014 $& $0.02228\,\pm0.00014$    & $0.02226\,\pm0.00014 $ & $0.02234\,\pm 0.00013$   \\
\hspace{1mm}\\

$\Omega_{\textrm{c}}h^2$& $0.1198\, \pm0.0025$& $0.1192\,\pm 0.0023$    & $0.1189\,\pm0.0024 $ &  $0.1204\,\pm 0.0024$   \\
\hspace{1mm}\\

$\tau$& $0.081\,\pm0.015$& $0.082\,^{+0.017}_{-0.015}$    & $0.064\,\pm 0.012$&   $0.080\,\pm 0.016$    \\
\hspace{1mm}\\

$10^{10}V_0/M^4$& $2.223\pm0.067$& $2.227\pm0.069$    & $2.141\,\pm 0.0048 $&   $2.223\,\pm0.069$   \\
\hspace{1mm}\\

$log(\sqrt{8\pi}b[GeV])$& $>19.5$& $>19.4$    & $>19.5$&   $20.4\,^{+1.4}_{-2.0}$    \\
\hspace{1mm}\\

$r$ &  $0.0029679\,_{-0.00000005}^{+0.0000014}$ &  $0.0029676\,_{-0.00000005}^{+0.0000021}$  & $0.0029679\,_{-0.00000001}^{+0.0000015}$&   $0.0029680\,_{-0.00000004}^{+0.0000013}$   \\
\hspace{1mm}\\

$N_{\rm eff}$ &  $3.06\pm0.10$&      $ 3.042\pm0.099$ & $ 3.032\,\pm0.099$   &  $ 3.111\,\pm 0.099$ \\
\hspace{1mm}\\

$H_0$ &      $67.46\pm0.52$&      $ 67.50\pm0.51$ & $ 67.51\,\pm0.51$   &     $ 67.88\,\pm 0.50$ \\
\hspace{1mm}\\

$\sigma_8$   & $ 0.833\,\pm0.013$   & $ 0.831\,\pm0.013$   & $ 0.8148\,\pm 0.0084$ &  $ 0.833\,\pm0.013$  \\
\hspace{1mm}\\

\hspace{1mm}\\
\hline
\hline

\end{tabular}}
\caption{$68 \% $ c.l. constraints on cosmological parameters in our extended $\Lambda$CDM+r+$N_{\rm eff}$ scenario from different combinations of datasets with a modified Starobinsky inflation.}
\label{table2}
\end{center}
\end{table}

\begin{figure}
\centering
\includegraphics[scale=1.0]{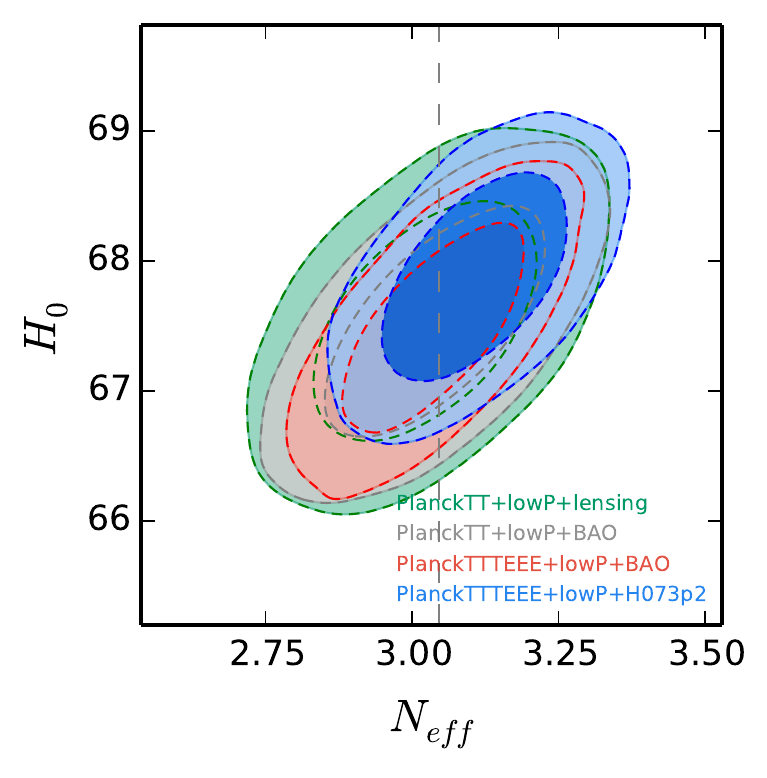}
\caption{Constraints at $68 \%$ and  $95 \%$ confidence levels on the $N_{\rm eff}$ vs $H_0$  plane, in our modified $\Lambda$CDM+r+$N_{\rm eff}$ Starobinsky scenario.}
\label{fignnu}
\end{figure}

Moreover, from Table~\ref{table2}, we notice that if an additional degree of freedom, for example a free dark radiation component is considered, then this theory produces a slightly better fit of the data by gaining about a $\Delta \bar \chi^2=2.9$, when considering PlanckTT+lowTEB and its combinations, with respect to a pure inflation $\Lambda$CDM+r+$N_{\rm eff}$ model. However, also in this case, the modified Starobinsky model has one less degree of freedom. We consider $N_{\rm eff}$ free to vary, because in this way data usually can accommodate inflationary potentials that are otherwise disfavored, thanks to its the degeneracy with the scalar spectral index $n_S$.

In this modified Starobinsky scenario, by introducing a dark radiation component free to vary $N_{\rm eff}$, we have very robust constraints for all the cosmological parameters, except $\Omega_ch^2$, which have no significant shifts with respect to $\Lambda$CDM+r model (see Table~\ref{table1}). In the minimal standard cosmological model (see Table~\ref{tablelcdm}), or in other inflationary models (see for example \cite{Tram:2016rcw,DiValentino:2016ucb}), to introduce a $N_{\rm eff}$ free to vary produces a value for this neutrino effective number higher than its expected value $3.045$ \cite{Mangano:2005cc,deSalas:2016ztq}, and a shift of all the parameters that are correlated with it. In particular, due to the strong correlation existing between $N_{\rm eff}$ and the Hubble constant $H_0$ (see Fig.~\ref{fignnu}), several authors have tried to solve the tension between the constraints coming from the Planck satellite \cite{planckparams2013} and \cite{planckparams2015} and the local measurements of the Hubble constant of Riess at al. \cite{R11} and \cite{R16},  thereby increasing the neutrino effective number \cite{darkradiation,edv1,ma1,DiValentino:2016ucb}. In the modified Starobinsky scenario, as showed for the standard Starobinsky model in \cite{Tram:2016rcw}, this solution is no longer suitable, because we have stronger bounds on the $N_{\rm eff}$, which is now in excellent agreement with its standard value, in all the cases. For example, if we look at the results for Planck TT+lowTEB, in the minimal standard cosmological model $\Lambda$CDM+r+$N_{\rm eff}$, we have $N_{\rm eff}=3.23^{+0.30}_{-0.36}$ and $H_0=68.9^{+2.7}_{-3.2}$ (see Table~\ref{tablelcdm}), while in the modified Starobinsky model for the same combination of data we find $N_{\rm eff}=3.06\pm0.13$ and $H_0=67.43\pm0.58$, with the error bars reduced at one third and one fifth with respect to the standard model. This happens because in the modified Starobinsky scenario, we are selecting a restricted number of models with a specific scalar spectral index, reducing the allowed parameter space for $N_{\rm eff}$ that is strongly degenerate with it. In this case, the tension on the Hubble constant becomes of $3.2\sigma$. Moreover, if we look at the value of $S_8=0.850\pm0.022$, the tension existing with KiDS-450 \cite{Hildebrandt:2016iqg} becomes of $2.3\sigma$. 

Although there is this inconsistency at more than $3 \sigma$ considering our scenario for the value of the Hubble constant and that one measured in \cite{R16}, we tried in any case to add the prior $H_0=73.2\pm1.7$ km/s/Mpc, to check the stability of our results. This prior produces a shift of half sigma of both $N_{\rm eff}=3.12\pm0.12$ and $H_0=68.00\pm0.56$ when adding H073p2 to Planck TT+lowTEB, but doesn't change our conclusions. Constraints are very robust and still prefer the standard value for the neutrino effective number and a Hubble constant in tension at about $2.8\sigma$ with \cite{R16}.

\begin{figure}
\centering
\includegraphics[scale=1.0]{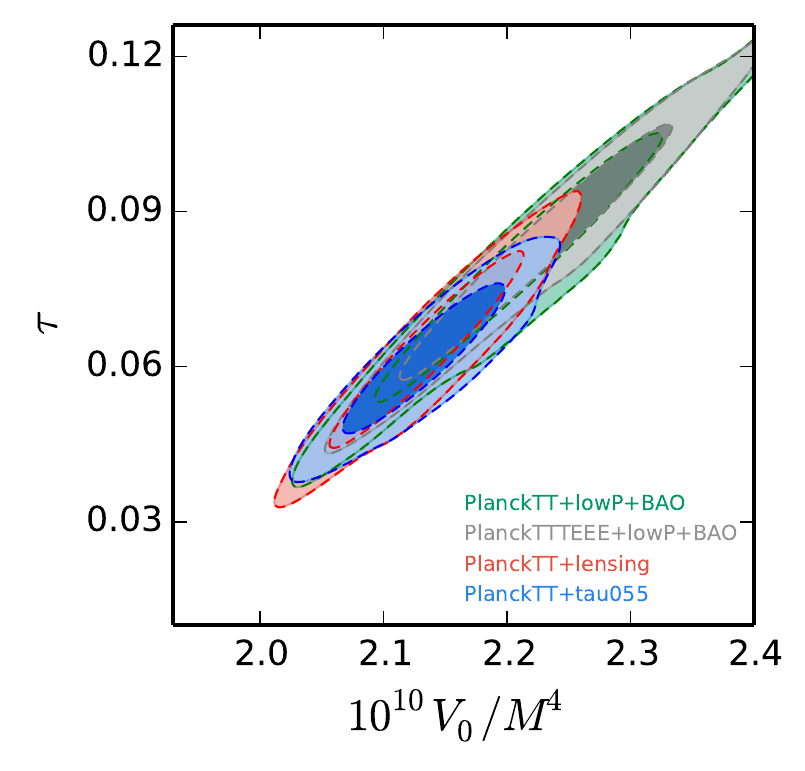}
\caption{Constraints at $68 \%$ and  $95 \%$ confidence levels on the $10^{10}V_0/M^4$ vs $\tau$  plane, in our modified $\Lambda$CDM+r+$N_{\rm eff}$ Starobinsky scenario.}
\label{figv0tau}
\end{figure}

Moreover, the cold dark matter density $\Omega_ch^2$ has very relaxed constraints by considering the $\Lambda$CDM+r+$N_{\rm eff}$ scenario (see Table~\ref{table2}), than the scenario without dark radiation (see Table~\ref{table1}). For example we have in the first scenario, that $\Omega_ch^2=0.1201\pm0.0033$ for Planck TT+lowTEB, with an uncertainty $3$ times larger than in the second case in which $\Omega_ch^2=0.1195\pm0.0012$. Also in this extended $\Lambda$CDM+r+$N_{\rm eff}$ scenario, as in the $\Lambda$CDM+r scenario, by adding the lensing dataset, the optical depth will shift towards a lower value becoming $\tau=0.064\pm0.012$ for Planck TTTEEE+lowTEB+lensing (see Fig.~\ref{figv0tau}).

Regarding the inflationary parameters in this extended $\Lambda$CDM+r+$N_{\rm eff}$ model, by considering the modified Starobinsky scenario, the same conclusions of the baseline $\Lambda$CDM+r are valid. We still predict a tensor-to-scalar ratio different from zero at several standard deviations, with a mean value of about $r\sim0.002968$. Moreover, $b$ still has a lower limit, i.e. $b > 1.2 \times10^7 GeV$ at $95 \%$ c.l. and $V_0=(2.209\pm0.072)\times 10^{10} M_P^4$ at $68\%$ c.l., for Planck TT+lowTEB datasets, slightly stronger than before. Certain combinations of datasets give us a bound at $1\sigma$ for $b$.

\begin{table}[!]
\begin{center}\footnotesize
\scalebox{0.65}{\begin{tabular}{lccccccc}
\hline \hline
       &   Planck TT  &  Planck TT &Planck TTTEEE  &Planck TTTEEE &Planck TTTEEE &Planck TTTEEE \\                     
        &        + lowTEB & + lowTEB + lensing    &  + lowTEB + BAO  & + lowTEB + lensing &+tau055&+ lowTEB + H073p2\\  
\hline
\hspace{1mm}\\

$\Omega_{\textrm{b}}h^2$& $0.02228\,\pm0.00019$    & $0.02224\,\pm0.00018 $& $0.02227\,\pm0.00013$    & $0.02227\,\pm0.00014 $&  $0.02227\,\pm 0.00014$ & $0.02227\,\pm 0.00013$   \\
\hspace{1mm}\\

$\Omega_{\textrm{c}}h^2$& $0.1195\,\pm 0.0012$    & $0.1188\,\pm0.0011 $& $0.11947\,\pm 0.00091$    & $0.11906\,\pm0.00093 $ & $0.11951\,\pm0.00094$ & $0.11950\,\pm 0.00091$   \\
\hspace{1mm}\\

$\tau$&  $0.076\,\pm0.017$    & $0.056\,\pm 0.013$& $0.081\,\pm0.015$    & $0.056\,\pm0.013$&  $0.0612\,\pm 0.0086$ & $0.079\,\pm 0.015$    \\
\hspace{1mm}\\

$10^{10}V_0/M^4$& $2.200\pm0.075$    & $2.104\,\pm 0.054 $& $2.220\pm0.067$    & $2.105\,^{+0.049}_{-0.058} $&  $2.136\,^{+0.038}_{-0.043}$ & $2.211\,\pm0.068$   \\
\hspace{1mm}\\

$log(\sqrt{8\pi}b[GeV])$& $20.4\,^{+1.9}_{-1.5}$    & $>19.6$&  $>19.6$    & $20.5\,\pm1.4$& $20.0\,^{+2.0}_{-1.0}$ & $20.4\,\pm1.5$    \\
\hspace{1mm}\\

$r$  &  $0.0029677\,_{-0.00000001}^{+0.0000018}$  & $0.0029681\,_{-0.000000003}^{+0.0000013}$ &    $0.0029681\,_{-0.000000001}^{+0.0000012}$  & $0.0029682\,_{-0.00000001}^{+0.0000010}$&   $0.0029676\,_{-0.00000005}^{+0.0000021}$& $0.0029677\,_{-0.00000003}^{+0.0000019}$   \\
\hspace{1mm}\\

$w$ &      $ -1.54\,^{+0.19}_{-0.39}$ & $ -1.40\,^{+0.27}_{-0.45}$   &      $ -1.025\,^{+0.058}_{-0.050}$ & $ -1.42\,^{+0.26}_{-0.45}$   &    $ -1.56\,^{+0.18}_{-0.38}$  &  $ -1.202\,\pm 0.061$ \\
\hspace{1mm}\\

$H_0$ &      $ >80.7$ & $ 81\,^{+20}_{-9}$  &      $ 68.2\,^{+1.4}_{-1.6}$ & $ 81\,^{+20}_{-7}$   &  $ >81.4$  &  $ 73.6\,\pm 1.9$ \\
\hspace{1mm}\\

$\sigma_8$      & $ 0.98\,^{+0.11}_{-0.06}$   & $ 0.92\,^{+0.12}_{-0.07}$ & $ 0.839\,\pm0.021$   & $ 0.92\,^{+0.12}_{-0.07}$ &   $ 0.97\,^{+0.10}_{-0.05}$ &  $ 0.887\,\pm0.022$  \\
\hspace{1mm}\\

\hspace{1mm}\\
\hline
\hline

\end{tabular}}
\caption{$68 \% $ c.l. constraints on cosmological parameters in our extended $w$CDM+r scenario from different combinations of datasets with a modified Starobinsky inflation.}
\label{table3}
\end{center}
\end{table}

\begin{figure}
\centering
\includegraphics[scale=1.0]{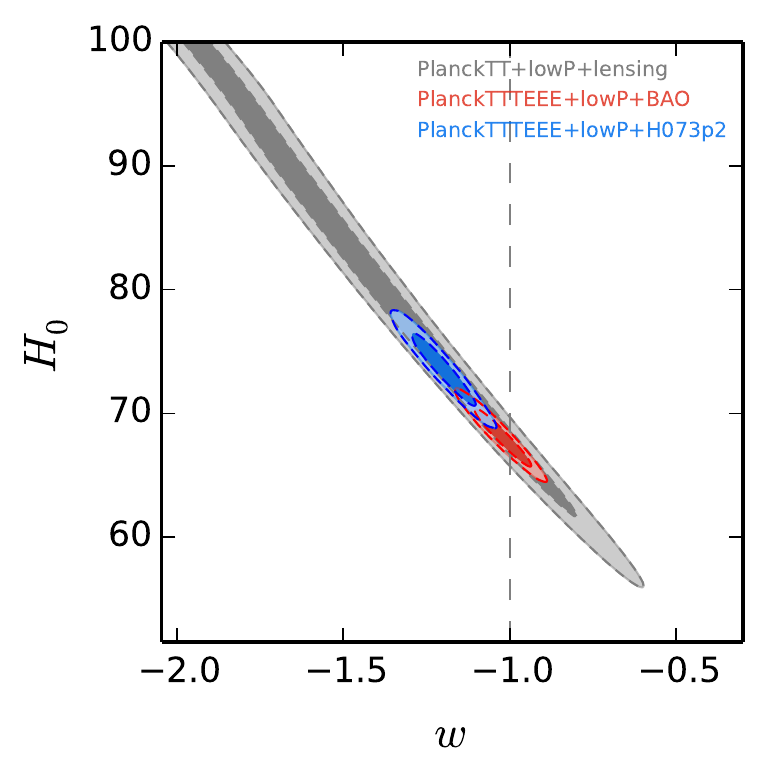}
\caption{Constraints at $68 \%$ and  $95 \%$ confidence levels on the $w$ vs $H_0$  plane, in our modified $w$CDM+r Starobinsky scenario.}
\label{figw}
\end{figure}

Afterward, we show the bounds for the $w$CDM+r scenario, using the Starobinsky inflationary model to test the modifications derived from quantum entanglement from this theory of the origin of the universe, in Table~\ref{table3}.

In our modified Starobinsky inflationary model, also by varying the equation of state of the dark energy, we find robust constraints for most of the cosmological parameters, which have no significant shifts with respect to $\Lambda$CDM+r model (see Table~\ref{table1}). However, due to the strong degeneracy existing between $w$ and $H_0$, see Fig.~\ref{figw}, we have a large shift of the Hubble constant. For this reason, one can try to consider a dark energy component free to vary, to solve the tension between the value of $H_0$ measured by the Planck satellite \cite{planckparams2013} and \cite{planckparams2015} and Riess at al. \cite{R11} and \cite{R16}. Also in this modified Starobinsky inflation, the tension can be solved if we consider an equation of state $w<-1$, as it has been shown by several authors (see for example \cite{DiValentino:2016hlg, Qing-Guo:2016ykt}).
In fact, we find that, when considering Planck TT+lowTEB+lensing, we have $w=-1.40\,^{+0.27}_{-0.45}$ and $H_0=81\,^{+20}_{-9}$ is in agreement within $1\sigma$ with \cite{R16}. In this case we can also solve the tension with the KiDS-450 experiment \cite{Hildebrandt:2016iqg}, finding $S_8=0.784\pm0.038$ perfectly in agreement with it.
By varying $w$, we can safely consider the prior H073p2 to Planck TTTEEE+lowTEB, and we find $w=-1.202\,\pm 0.061$, smaller than the cosmological constant by more than $3\sigma$ and $H_0=73.6\,\pm 1.9$. However, in this case, we restore the tension at $2\sigma$ level with the KiDS-450 experiment \cite{Hildebrandt:2016iqg}, obtaining $S_8=0.830\pm0.019$.

Moreover, when we consider the BAO dataset, the dark energy equation of state is consistent with a cosmological constant with equation of state $w=-1$, thus a slight tension at $2\sigma$ reappears between this Hubble constant and \cite{R16}.

Regarding the inflationary parameters, we find the same results of the previous scenario $\Lambda$CDM+r. We have the same predicted tensor-to-scalar ratio. The 'SUSY-breaking' scale associated with the landscape effects $b$ has a lower limit, i.e. $b>1.3\times10^7 GeV$ at $95 \%$ c.l. and $V_0=(2.200\pm0.075)\times 10^{10} M_P^4$ at $68\%$ c.l., for Planck TT+lowTEB. Again we find that certain combinations of datasets give an indication at $1\sigma$ for $b$, as for example Planck TTTEEE+lowTEB+lensing.

\begin{table}[!]
\begin{center}\footnotesize
\scalebox{0.8}{\begin{tabular}{lcccc}
\hline \hline
    &Planck TT  &  Planck TTTEEE&Planck TTTEEE\\                     
         & + lowTEB &      + lowTEB    &   + lowTEB + lensing \\  
\hline
\hspace{1mm}\\

$\Omega_{\textrm{b}}h^2$& $0.02229\,\pm 0.00019 $&  $0.02229\,\pm 0.00014$& $0.02226\,\pm 0.00014$   \\
\hspace{1mm}\\

$\Omega_{\textrm{c}}h^2$& $0.1195\, \pm0.0032$&  $0.1196\,\pm0.0025$ & $0.1185\,\pm 0.0024$   \\
\hspace{1mm}\\

$\tau$& $0.076\,\pm0.017$& $0.077\,\pm 0.016$& $0.058\,\pm 0.013$    \\
\hspace{1mm}\\

$10^{10}V_0/M^4$& $2.195\pm0.074$&  $2.201\,\pm 0.068$&  $2.109\,\pm0.052$   \\
\hspace{1mm}\\

$log(\sqrt{8\pi}b[GeV])$& $>19.5$&  $20.6\,\pm1.4$& $20.6\,\pm1.4$    \\
\hspace{1mm}\\

$r$ &  $0.0029679\,_{-0.00000001}^{+0.0000015}$ &  $0.00296836\,_{-0.00000002}^{+0.00000075}$ &   $0.00296831\,_{-0.00000002}^{+0.00000085}$   \\
\hspace{1mm}\\

$N_{\rm eff}$ &  $3.05\pm0.12$&   $ 3.05\,\pm 0.10$ &    $ 3.023\,\pm 0.099$ \\
\hspace{1mm}\\

$w$ &  $-1.53\,^{+0.19}_{-0.41}$&  $ -1.55\,^{+0.18}_{-0.38}$ &  $ -1.41\,^{+0.27}_{-0.45}$ \\
\hspace{1mm}\\

$H_0$ &      $>80.5$&   $ >81.6$ &   $ 81\,^{+20}_{-9}$ \\
\hspace{1mm}\\

$\sigma_8$   & $ 0.975\,^{+0.11}_{-0.06}$ &  $ 0.99\,^{+0.10}_{-0.05}$ &   $ 0.92\,^{+0.12}_{-0.07}$  \\
\hspace{1mm}\\

\hspace{1mm}\\
\hline
\hline

\end{tabular}}
\caption{$68 \% $ c.l. constraints on cosmological parameters in our extended $w$CDM+r+$N_{\rm eff}$ scenario from different combinations of datasets with a modified Starobinsky inflation.}
\label{table4}
\end{center}
\end{table}

Finally, In Table~\ref{table4} we show the bounds for the $w$CDM+r+$N_{\rm eff}$ scenario for this theory.

Again we find no significant shifts with respect to $\Lambda$CDM+r model (see Table~\ref{table1}), and the extended parameters $N_{\rm eff}$ and $w$ give the same results as in the Tables~\ref{table2} and \ref{table3}. Our previous conclusions are not affected by the introduction of more degrees of freedom in the model. The implication coming out of these tests is a strong bound for the neutrino number $N_{\rm eff}$, in excellent agreement with its standard value, and $w<-1$ and $H_0$ that is in agreement within $1\sigma$ with \cite{R16} for Planck TTTEEE+lowTEB+lensing.

\subsection{The Quantum Landscape Multiverse}

A theory for the origin of the universe from a quantum landscape multiverse was proposed and developed in \cite{archillmh, richlmh, tomolmh, lmh}. By studying the wavefunction of the universe through the landscape by means of quantum cosmology, this proposal was able to explain why the most probable initial state of the universe is at high energies. Decoherence and the entanglement of our branch of the wavefunction with all others led to a series of predictions derived in \cite{tomolmh}. The effect comes from the quantum entanglement of our universe with all others and it imprints a series of anomalies in the observables of our sky. However the initial derivation of the modifications from entanglement, imprinted in the CMB and the large scale structure of the universe, was illustrated in \cite{tomolmh} for a convex inflationary potential, the exponential model. 

The calculation of the modification from entanglement to the inflaton potential, Einstein equations, the gravitational potential and power spectrum, were recently extended to the class of concave downwards potentials in \cite{lmh}. The class of concave downwards potentials involves subtleties and need careful derivations, whenever the semiclassical approximation is invoked in quantum cosmology, which was the case for implementing the calculations in this theory. We now know that concave downwards potentials, (with the Starobinsky model and the hilltop inflation models as representatives), are the potentials favored by the Planck collaboration data. The predictions for modifications in the CMB and LSS recently derived for concave potentials in \cite{lmh}, are based on nonlocal modifications that entanglement introduces to the field potential $V_{eff}$. As reviewed in Section \ref{sec:model}, these modify the Friedmann equation, the field solutions $\phi[k]$ and consequently the gravitational potential of the universe  $\Phi_{0}$ by an amount $\delta \Phi$, $\Phi \simeq \Phi_{0} + \delta \Phi$. Since the gravitational potential is related to the energy density, then the effective potential correction can be related to a gravitational potential modification, through the Poisson equation, $\nabla^{2} \delta \Phi = 4\pi f(b,V)$. Note that the modification term $f[b,V]$ is negative. 

The modified gravitational potential of the universe, and an upper bound for $b$ from the flatness condition on the inflaton potential, were discussed in detail in \cite{tomolmh}. Here the lower bound on $b$ is obtained from analyzing the Planck data. 

\begin{figure}
\centering
\includegraphics[width=7.4cm]{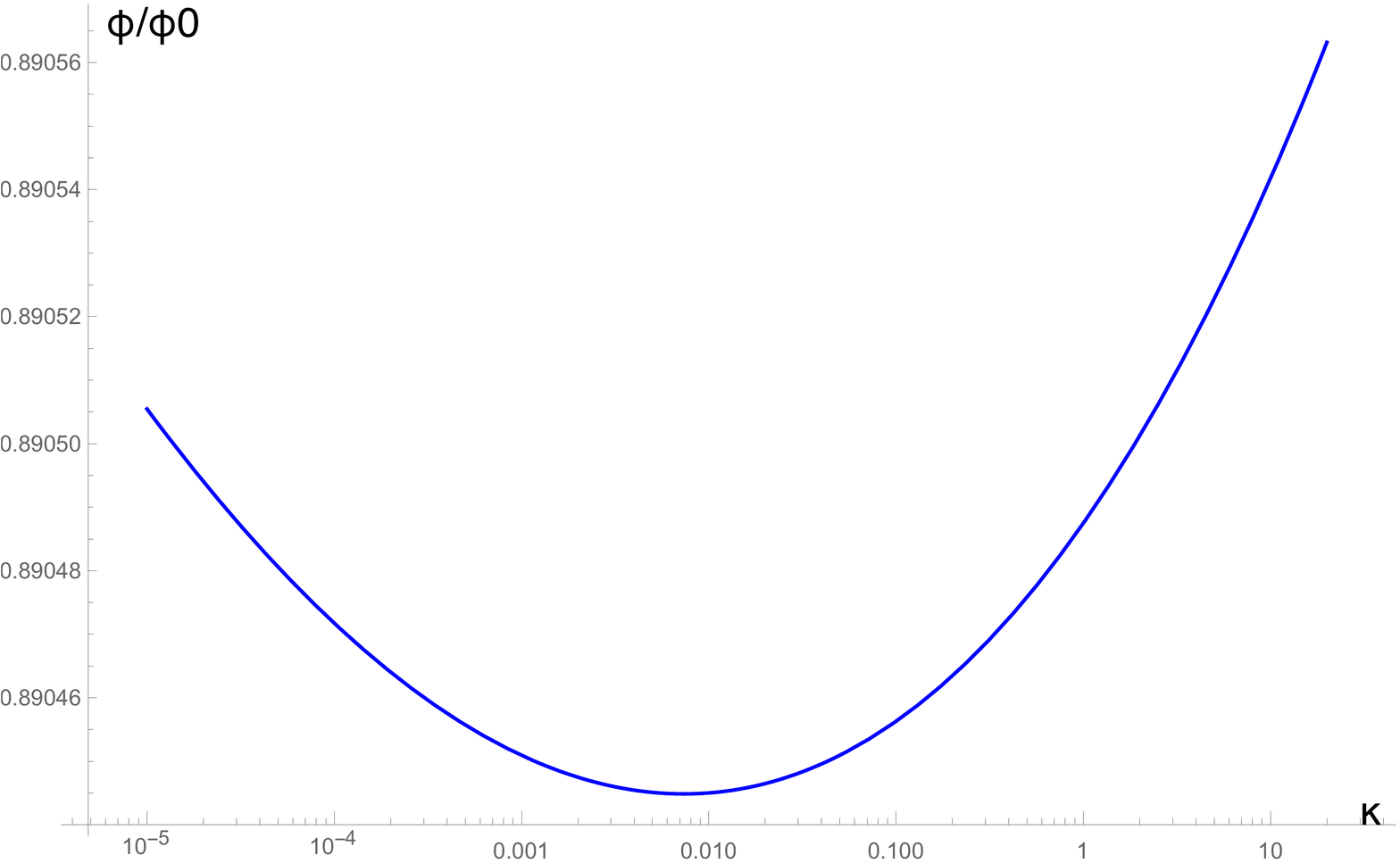}
\includegraphics[width=7.4cm]{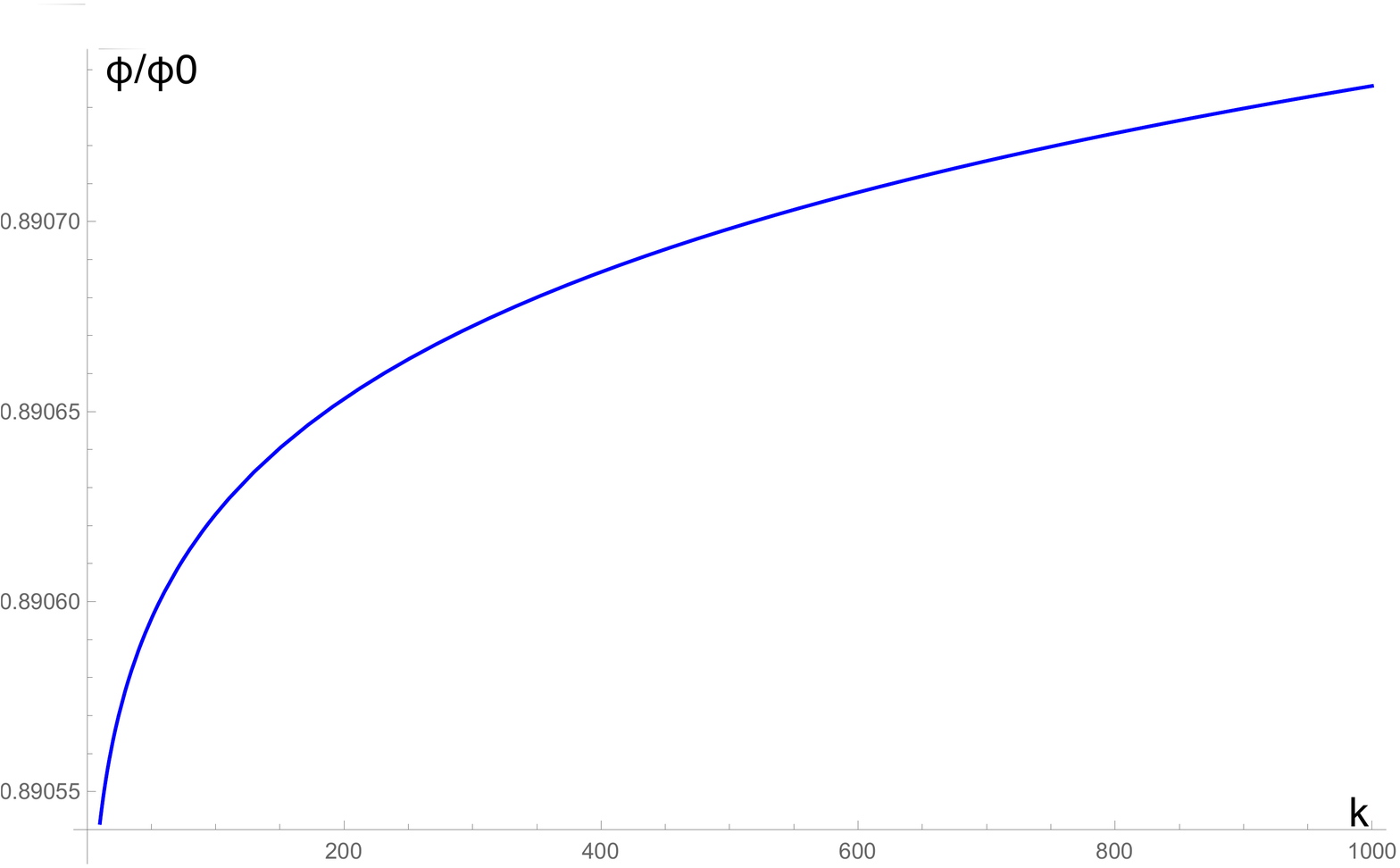}
\includegraphics[width=7.4cm]{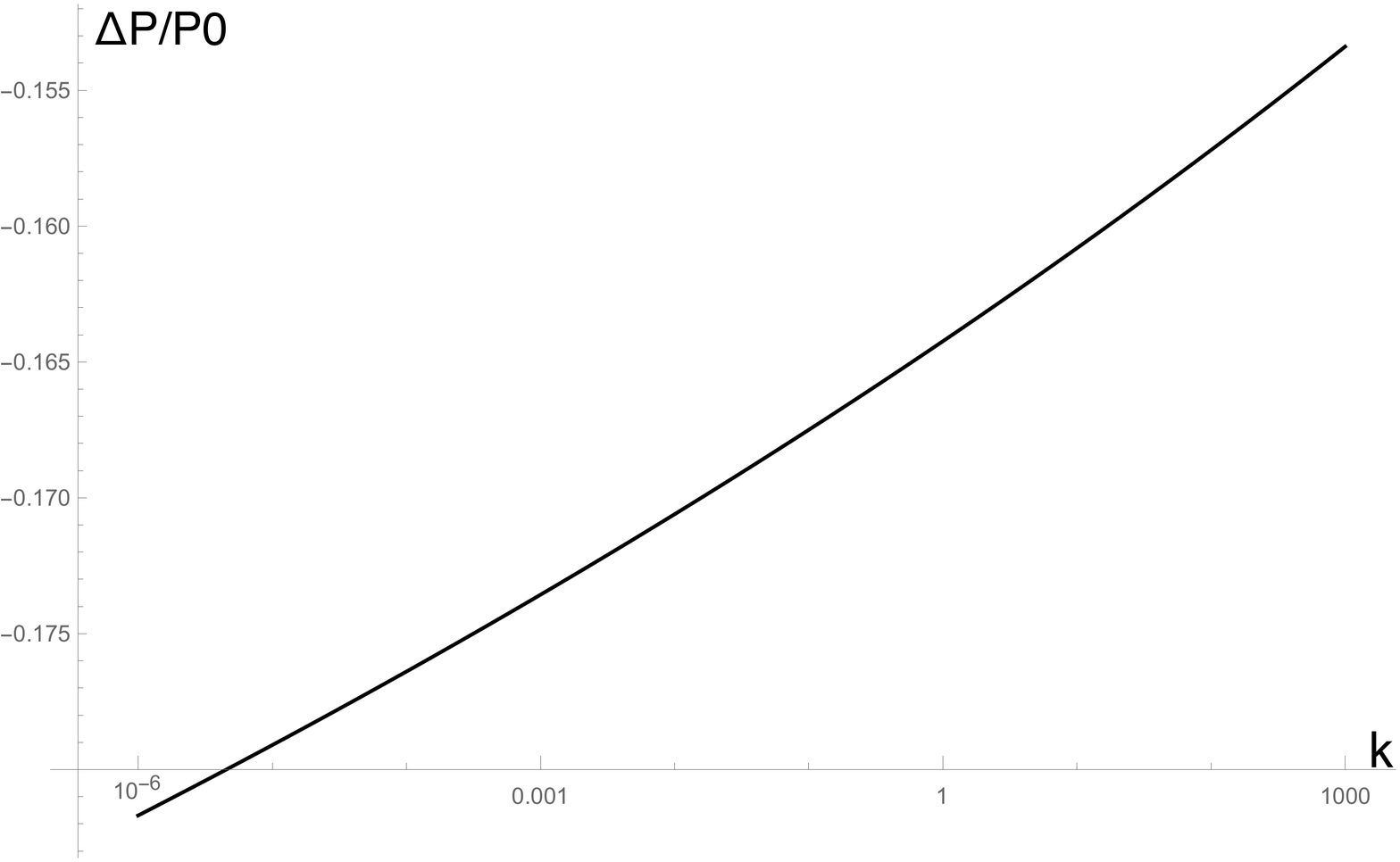}
\caption{The ratio of the modified field solution $\phi[k]/\phi_{0}[k]$ and the modified power spectrum $\frac{P[k] - P_{0}[k]}{P_{0}[k]}$ for a value of the SUSY-breaking scale $b=1.56 \times 10^7 GeV$ and $V_{0} =2.25 \times 10^{-10} M_{p}^4$ within the 2 sigma allowed region in the likelihood plots of Section \ref{results}. We chose this value of b just as an illustration of the effects, since the difference in the power spectrum is more dramatic. It can be seen that the power is suppressed by about $14\%$ at low multipoles producing a suppression of temperature autocorrelation at low multipoles, and the field changes behaviour at the dipole level $k \approx (0-1) hMpc^{-1}$ and at $k \approx 20 hMpc^{-1}$ leading to suppressed perturbations of these $k-$modes, corresponding to the power asymmetry, and a giant void at $z\sim1$ of size about $200 Mpch^{-1}$. The overall behaviour persists for different values of $b,V$. However the effect gets smaller as $b$ increases.}
\label{starofig}
\end{figure}

The key to our analysis are finding the bounds on the parameter $b$ which control the modification term. If the allowed value by current data on $b$ is such that the correction $f[b, V]$ to the potential $V_{eff}$ is $ 10 - 20\% $ of the unmodified potential $V(\phi)$ then the series of predictions for the anomalies made in \cite{tomolmh} for this theory, fall within the observed range. Explicitly, analyzing the modifications against current data for the model at hand, we find the following: the slow roll of the field changes at the lowest $k \simeq (0-1) hMpc^{-1}$ and at around $k\simeq 20 hMpc^{-1}$ leading to suppressed perturbations at those modes and efolds, see Fig.~\ref{starofig}. These suppressed perturbations derived from the modified solution of the slow rolling field and the modified potential, can also be seen in the modification in the gravitational potential correction $\delta \Phi[k]$. The $k ~ O(1)$ suppression produces a power asymmetry at the dipole level. Using the standard relations to translate these modes numbers into sizes and redshifts, as was done in detail in \cite{tomolmh} ($k \simeq 1/a hMpc^{-1}$ where $a$ is the scale factor), then the suppression at $k\simeq 20 hMpc^{-1}$ corresponds to a density suppression by about $30\%$ for the lower bound $b \simeq 1.5 \times 10^{7}GeV$, which is most easily seen from the gravitational potential as a void. The $ k\simeq 20 hMpc^{-1}$ corresponds to a present day redshift of about $z \simeq 1$ and size of about $200 Mpch^{-1}$. Physically, the first of the suppressed perturbation modes $k\simeq 1 hMpc^{-1}$ occurs because the modification term is strongest at the start of slow roll, i.e. at the lowest $k$ and gradually redshifts away. The second disturbance to the field's slow roll, producing suppressed perturbations at that efold, occurs at about $k\simeq 20 hMpc^{-1}$ and produces a giant void. This discontinuity is due to the interplay of the strength of the $Log$ versus the $Exponential$ terms (having different signs, see \cite{tomolmh}) in the modification term in the potential $f[b,V]$. However the modification of the potential $V_{eff} - V = f[b,V]$ which depends on the SUSY  breaking scale $b$ should be at least of order $10\%$  of the unmodified potential, in order to produce a significant effect in the suppression of the power spectrum and the correct density contrast suppression for the void. To avoid repetition, we refer the reader to \cite{archillmh, richlmh, tomolmh, lmh} for detailed derivations of the effects summarized here. 

By placing lower bounds on $b$ through the data, we find that all the anomalies, including the giant void, power asymmetry and suppression of the temperature spectrum at low multipoles, are produced for $1.2 \times 10^{7} < b < 1 \times 10^{8} GeV$, while being in excellent agreement with all the astrophysical data accumulated so far. The overall features of the anomalies, such as the suppression of power and suppressed perturbation modes, remains for all values of $b$. However the effect becomes less significant as $b$ is dialed up to higher values because the correction terms become really small relative to the unmodified potential. For our case that value of $b$ where the modification term is less than $5\%$ of the potential is $b=6\times 10^{8} GeV$.

We here derived the CMB and LSS observables, such as the power spectra, the scalar spectral index, the gravitational potential, and the tensor-to-scalar ratio for the modified Starobinsky model obtained from this theory, and analyzed it against the most recent available datasets, including the polarization data. The analysis shows that the modified Starobinsky model can fit the Planck data as well as the minimal standard cosmological model $\Lambda$CDM+r. Further, the constraint we find on the parameter from the most recent data available is $b>1.2 \times 10^{7} GeV$ which lie within the $2 \sigma$ region. This is our most important finding for the concave type potentials: $b$ and $f[b,V]$ from the data constraints, allow for modifications to the potential to be in the range $V_{eff}$ be $20\% - 30\%$ in strength at the onset of inflation. This finding is significant because it confirms that the allowed range for $b$ and $V$ are such that these models are in excellent agreement with Planck temperature and polarization datasets, while being sufficient to explain the series of anomalies, first predicted in \cite{tomolmh} for this theory, namely: a power suppression at low multipoles, which for the lower limit on $b$ we find here corresponds to a power suppressed by $12 \%$ at low-$\ell$'s; a modification strength $f[b,V]$ to the inflaton potential by $30\%$ leading to a disturbance of the slow rolling behaviour of perturbations at very low k, (those around the dipole/quadrupole level), exhibiting as a power asymmetry between the the two hemispheres; another discontinuity at around $k\simeq 20 hMpc^{-1}$ showing today as a giant void at $z\simeq 1$ and size of about $\simeq 200 Mpch^{-1}$; a slightly suppressed overall $r[k]$; a very mild running for the scalar index $n[k]$; and $\sigma_8 \simeq 0.8$.

\section{Conclusions}
\label{sec:conclusions}

In conclusion, we find that the Starobinsky model, modified by a nonlocal corrections term, originating from the entanglement of our branch of the wavefunction with all else in the quantum landscape multiverse, not only falls within the allowed range of the Planck data on CMB temperature and polarization, but it is also consistent with the range for $b$ that allows the landscape multiverse to explain a serie of anomalies, such as power suppression and the giant void, present in the current Planck 2015 data \cite{planck1,planck2,planck2013}.

The lower bound $b>5.3 \times 10^7 GeV$ and the constraint $V_{0}=2.209 \times 10^{-10} M_{p}^4$ at $1\sigma$ level are obtained here from analyzing the Planck 2015 data. We find that the modified Starobinsky model can fit the Planck data as well as the minimal standard cosmological model $\Lambda$CDM+r, by improving the $\Delta \bar \chi^2=2.5$ for Planck TT+lowTEB data, probably because it can better recover the first peak and the lower quadrupole of the temperature angular power spectrum. Anyway, in the modified Starobinsky model, we have one less degree of freedom, because $r$ is a derived parameter and we are varying only $6$ parameters instead of $7$, as in the minimal standard cosmological model $\Lambda$CDM+r. Interestingly, for some combination of datasets, as for example Planck TT+lowTEB+lensing datasets, some indication at $1\sigma$ level of an upper and lower bound appear for the 'SUSY breaking' scale $b$,  $1.2\times10^7GeV<b<6.5\times10^8GeV$. Moreover, we have a very precise and robust expectation for the derived value of the tensor-to-scalar ratio $r\sim0.002968$. All the other cosmological parameters have no significant shift with respect to the minimal standard cosmological model $\Lambda$CDM+r.
Moreover, we tested the robustness of our results and, also considered extensions of the number of parameters that we are letting free to vary. First, we consider the effective number of relativistic degrees of freedom $N_{\rm eff}$ because it is strongly degenerate with the scalar spectral index $n_S$, and with this additional degree of freedom the data can usually accommodate inflationary potentials that are otherwise disfavored. We find instead, that in our modified Starobinsky scenario, the number of neutrino species is perfectly in agreement with the theoretical expected value $N_{\rm eff} =3.045$. We tried also to consider a dark energy equation of state $w$ free to vary, because in this way we can solve the tensions between Planck and external datasets, as the direct measurements of the Hubble constant \cite{R16} and the $S_8$ \cite{Heymans:2012gg,Hildebrandt:2016iqg}. In our modified Starobinsky scenario, we find that, also when considering Planck TT+lowTEB+lensing, a $w<-1$ can solve both the tensions. All the other cosmological parameters, in particular $b$, have no significant shift with respect to baseline $\Lambda$CDM+r modified for the Starobinsky model.

Finding a model, which theoretically tells a consistent and coherent story of the evolution of the universe from its quantum infancy before it underwent inflation to present, is obviously important. But if that model goes beyond a theoretical explanation of the origin of the universe and it fits the standard cosmological parameters such as the temperature and polarization spectra, and predicts a series of anomalies in agreement with the Planck experiment, then it becomes more exciting since it offers us new hints into the physics of the universe from the time the universe was a quantum wavefunction and beyond.

Of course it remains to be seen whether these findings remain solid for other models of inflation. Since the class of inflationary models is large, we can not analyze all of them one by one in the light of this theory. Instead in a companion paper we will analyze the class of hilltop models of inflation to complete the study of concave downwards potentials, and revisit the exponential model \cite{DiValentino:2016ziq}.

{\it Note on the History of The Model}

The model of inflation discussed here is mostly known in literature as the Starobinsky model. The history of this model was brought to our attention by various colleagues. Although the history of the model is not related to the scope of the paper, we thought it is of interest and important to share it here. In 1980 Starobinsky proposed the $R^2$ model similar to the current one in the context of Einstein gravity with conformal anomaly and his focus was on particle creation, in particular graviton production.  Since this paper appeared before the theory of inflation was proposed, it was not focused in addressing the usual issues that flatness, homogeniety and isotropy, that inflationary models address. In 1981 Mukhanov and Chibisov studied perturbations of the model and in 1983 Starobinsky  introduced the current $R + R^2$ version of the model, also studied independently in the same year by Barrow and Ottewill. In 1984 Whitt showed that by changing from Einstein to Jordan frame this model can be equivalently described by a scalar field rolling down a potential V as the one in Eq.4. Then in 1985, Kofman, Linde and Starobinsky, modified the proposal of the 1980 paper by Starobinsky, and explicitly showed the modified version $R + R^2$ to be a model of inflation in the plethora of the inflationary paradigm models. Further papers in 1988 by Barrow, and Maeda, discussed this model as an inflationary model. The papers mentioned here in an historic context can be found below in Ref. \cite{staro,Kofman:1985aw, Goncharov:1983mw, Starobinsky:1982mr, Whitt:1984pd, Barrow:1988xi, Barrow:1988xh, Maeda:1987xf, Barrow:1983rx}. Since then, there is a vast amount of literature investigating various aspect of this model but our purpose in this note was to provide its history.

\acknowledgments
We would like to thank F. R. Bouchet,  A. Melchiorri, A. Linde and J. Barrow, for stimulating discussions and comments, and the Planck Editorial Bord for taking the time to review the paper and to approve it. This work has been done within the Labex ILP (reference ANR-10-LABX-63) part of the Idex SUPER, and received financial state aid managed by the Agence Nationale de la Recherche, as part of the programme Investissements d'avenir under the reference ANR-11-IDEX-0004-02. LMH acknowledges support from the Bahnson funds. 

\begin{table}[!]
\begin{center}\footnotesize
\scalebox{1.0}{\begin{tabular}{lccccc}
\hline \hline
          &Planck TT & Planck TTTEEE&Planck TT& Planck TTTEEE\\                     
         & + lowTEB     &        + lowTEB   & + lowTEB       &  + lowTEB \\  
\hline
\hspace{1mm}\\

$\Omega_{\textrm{b}}h^2$&  $0.02224\,\pm0.00023 $& $0.02225\,\pm 0.00016$& $0.02240\,\pm 0.00037$ & $0.02219\,\pm 0.00025$   \\
\hspace{1mm}\\

$\Omega_{\textrm{c}}h^2$&  $0.1195\,\pm0.0022 $& $0.1197\,\pm0.0014$ & $0.1214\,^{+0.0039}_{-0.0043}$ & $0.1191\,\pm 0.0031$   \\
\hspace{1mm}\\

$\tau$&  $0.077\,\pm 0.019$& $0.078\,\pm 0.017$& $0.082\,^{+0.021}_{-0.025}$ & $0.076\,\pm 0.018$    \\
\hspace{1mm}\\

$log(10^{10}A_S)$&  $3.087\,\pm 0.036 $& $3.092\,\pm 0.033$& $3.102\,^{+0.046}_{-0.052}$ & $3.086\,\pm0.038$   \\
\hspace{1mm}\\

$n_S$&  $0.9666\,\pm0.0062$& $0.9652\,\pm0.0047$& $0.974\,\pm0.016$ & $0.963\,\pm0.010$    \\
\hspace{1mm}\\

$r$ & $<0.0472$&  $<0.0463$ &  $<0.0529$& $<0.0487$   \\
\hspace{1mm}\\

$N_{\rm eff}$ &   $ (3.046)$   &  $ (3.046)$ &  $ 3.23\,^{+0.30}_{-0.36}$  &  $ 3.00\,\pm0.21$ \\
\hspace{1mm}\\

$H_0$ &       $ 67.42\,\pm0.99$   &  $ 67.31\,\pm 0.64$ &  $ 68.9\,^{+2.7}_{-3.2}$  &  $ 66.9\,\pm 1.7$ \\
\hspace{1mm}\\

$\sigma_8$      & $ 0.828\,\pm 0.014$ &  $ 0.830\,\pm0.013$ &  $ 0.837\,^{+0.022}_{-0.025}$ &  $ 0.827\,\pm0.018$  \\
\hspace{1mm}\\
\hline \hline
\end{tabular}}
\caption{$68 \% $ c.l. constraints on cosmological parameters considering the minimal standard cosmological model for different combinations of datasets.}
\label{tablelcdm}
\end{center}
\end{table}





\end{document}